\documentclass{mn2e}

\usepackage{epsfig} 
\usepackage{graphicx}
\usepackage{amsmath}
\usepackage{natbib}
\usepackage{subfigure}
\usepackage[T1]{fontenc}
\usepackage{aecompl}

\begin{document}

\title[ESO148-IG002]{The energy source and dynamics of infrared luminous galaxy
ESO 148-IG002}

\author[S. Leslie et al.]{Sarah K. Leslie$^1$\thanks{E-mail:
u5022102@anu.edu.au}, Jeffrey A. Rich$^2$, Lisa J. Kewley$^{1,3}$, Michael A.
Dopita$^{1,3,4}$,\\
$^1$Research School of Astronomy and Astrophysics, Australian National
University, Cotter Road, Weston, ACT 2611, Australia\\
$^2$Carnegie Observatories, 813 Santa Barbara Street, Pasadena, California,
91101 USA\\
$^3$Institute for Astronomy, University of Hawaii, 2680 Woodlawn Drive,
Honolulu, HI 96822\\
$^4$Astronomy Department, Faculty of Science, King Abdulaziz University, PO Box
80203, Jeddah, Saudi Arabia}
\maketitle



\begin{abstract}
ESO 148-IG002 represents a transformative stage of galaxy evolution, containing
two galaxies at close separation which are currently coalescing into a single
galaxy. We present integral field data of this galaxy from the ANU Wide Field
Spectrograph (WiFeS). We analyse our integral field data using optical line
ratio maps and velocity maps. We apply active galactic nucleus (AGN), star-burst
and shock models to investigate the relative contribution from star-formation,
shock excitation and AGN activity to the optical emission in this key merger
stage. We find that ESO 148-IG002 has a flat metallicity gradient, consistent
with a recent gas inflow. We separate the line emission maps into a
star forming region with low velocity dispersion that spatially covers the whole
system as well as a southern high velocity dispersion region with a coherent velocity pattern which could either be rotation or an AGN-driven outflow, showing little evidence for pure star formation. We show that the two overlapping galaxies can be
separated using kinematic information, demonstrating the power of moderate
spectral resolution integral field spectroscopy.

\end{abstract}


\section{Introduction}\label{Sec:intro}

Understanding the processes involved in galaxy formation and development is a
pressing problem in modern astrophysics. Understanding galaxy mergers is key to
understanding the formation of elliptical galaxies \citep{Toomre1977}.

Ultraluminous ($L_{IR}>10^{12}L_\odot$) and Luminous ($L_{IR}>10^{10}L_\odot$) Infrared Galaxies (U/LIRGs) emit more energy in
infrared (5-1000$\mu$m) than at all other wavelengths combined 
\citep{Sanders1996}. U/LIRGs are more common at higher redshifts than locally and
by z=1 they form the dominant component of the IR luminosity function
(\citealt{Elbaz2002}). The majority of LIRGs are formed by strong
interactions/mergers of gas rich spirals \citep{Sanders1988}. The enormous gas concentrations
involved facilitate phenomena such as powerful starbursts with accompanying
galactic winds, and the feeding of Active Galactic Nuclei (AGNs) which could
contribute to the infrared luminosity \citep{Sanders1996,Lonsdale2006}. Resolving the detailed
ionization structure, kinematics and power sources of local U/LIRGs will further
our understanding of galaxy evolution both locally and at higher redshift 
\citep{Martin2005,Rupke2005,Rich2011}.

As galaxies collide, large quantities of the gas in the disk of each galaxy
propagate towards the central regions \citep{Barnes1996}. Gravitational forces
between the two galaxies produce tidal tails and disrupted morphology. Tidally
induced gas motions and outflows from galactic winds become increasingly common
as the merger progresses (e.g., \citealt{Hopkins2013}). Shocks induced by such
large-scale gas flows can influence the emission line gas
\citep{Armus1989,Heckman1990,Colina2005,Zakamska2010}. This may contaminate line ratios used to
determine metallicity, star formation rate (SFR), and power source
\citep{Rich2011}.

\citet{Yuan2010} explains how the composite merger is a critical stage for
studying the spectral evolution of two galaxies coalescing. In particular,
composite galaxies which are in the process of forming one system, form a bridge
between pure starburst and Seyfert galaxies.
Shock excitation and its effects on emission line spectra are complex; for
example, shock excitation can exhibit extended low-ionization narrow
emission-line region (LINER) emission characteristics, or shocks can mimic starburst-AGN composite activity (e.g \citealt{Rich2010,Rich2011}). Integral field spectroscopy (IFS) is a
powerful tool for separating various spatial and kinematic components associated
with different power sources (for example \citealt{Monreal-Ibero2006, Harrison2012,Cano-Diaz2012,Harrison2014,FoersterSchreiber2014}).
In this paper we investigate the power sources in ESO 148-IG002. ESO 148-IG002 has a redshift of z=0.045, and is
is a late stage ULIRG merger between two disk galaxies of similar size with
$L_{IR} \cong 2 \times10^{12}L_{\odot}$ \citep{Johansson1988}. The system still contains
the nuclei of the original galaxies separated by 5 kpc in projection, and there is evidence that the bright southern nucleus contains an AGN based on [NeV], [OIV], and [NeII] fine-structure emission line diagnostics
\citep{Petric2011} and X-ray colour criterion \citep{Iwasawa2011}.

We discuss the observations and data in Section 2. In Section 3, we present the
results of our observations by analysing line ratio maps and emission line
diagnostic diagrams. We use the distribution of velocity dispersions to separate
the galaxy's ionizing sources and discuss rotation maps in Section 4. In Section
5 we employ shock, AGN and HII region models to determine the roles that shocks
and AGN activity play in this system. We also investigate the metallicity
properties of this galaxy. In Section 6 we discuss the implications of our
results. Finally, we give our conclusions in Section 7.

Throughout this paper we adopt the cosmological parameters
$H_0=70$ km s$ ^{-1}$Mpc$ ^{-1}$,
$\Omega_\lambda=0.72$, and  $\Omega_M=0.28$, based on the five-year WMAP results
\citep{Hinshaw2009} and consistent with the \citet{Armus2009} summary of the GOALS
sample. With this cosmology at the redshift of ESO 148-IG002, 1" corresponds to 880~pc.

\begin{figure}
\centerline{\includegraphics[scale=0.4]{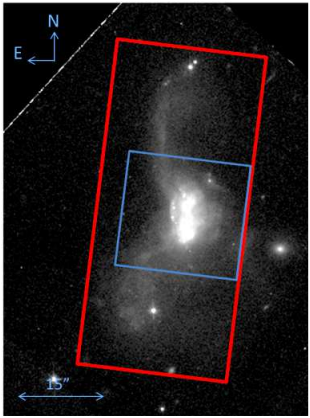}}
\caption{Hubble Space Telescope broad I band image of ESO 148-IG002 taken with the F814w filter on the WFPC2 instrument. The
combined field of view of our two WiFeS pointings is overlaid in red and 15" shown for
scale. The blue rectangle shows the region with the sufficiently high
signal-to-noise for our analyses, and hence only this section is shown in
subsequent maps.}\label{aladin}
\end{figure}

\section{ESO 148-IG002; Observations, data reduction and line
fitting}\label{Sec:obs}
\subsection{ESO 148-IG002 (IRAS 23128-5919)}
In appearance, this strongly interacting system is very similar to the nearby
NGC 4038/4039 (`Antennae') pair.
Previous studies of ESO 148-IG002 (with \textit{IRAS} designation IRAS
23128-5919) show that the galaxy has two nuclei separated by $\sim 4.5$"
\citep{Zenner1993, Duc1997}, surrounded by bright, possibly star forming knots. The
system has two faint tidal tails curling in opposite directions, indicating
there was originally two dynamically independent galaxies.
\cite{Bergvall1985}, and \cite{Johansson1988} found LINER
and starburst features in the nuclear spectra. They suggested that the nuclear
emission-line ratios are consistent with shocks associated with a star burst
driven galactic wind. Further studies of infrared mergers by \citet{Lipari2003},
agree with this interpretation.
Wolf-Rayet (WR) features have also been detected in
the southern nucleus by  \citet{Lipari2003}. \cite{Johansson1988} found
a broad intensity enhancement around $\lambda = 4650$ \AA\ which they interpret
as NIII $\lambda 4641$ and HeII $\lambda 4686$ emission from WR stars of type N.
The nucleus of the southern galaxy, although radio quiet, is relatively bright
and pointlike, and is suggested to host an active nucleus \citep{Petric2011,Iwasawa2011,Bushouse2002,Stierwalt2013}.

VLT-SINFONI integral field spectroscopic observations have been taken of ESO 138-IG002 by \cite{PiquerasLopez2012} in the $H$(1.45-1.85$\mu$m) and $K$(1.95-2.45$\mu$m) bands, covering the central $\sim$7-11 kpc. From Pa$\alpha$ equivalent width measurements, they find that the northern nucleus is dominated by star formation, whereas in the southern nucleus, the presence of an AGN is supported by strong compact [SiVI] emission. We compare our kinematic results with the findings of \cite{PiquerasLopez2012} in Section 4. \cite{Rodriguez-Zaurin2011} use the VLT/VIMOS instrument to study H$\alpha$ in LIRGS and ULIRGS. They are unable to detect ESO 148-IG002's tidal tails in the ionized gas emission and report that the optical spectrum shows a mix between LINER, Sy2 and HII-like features.

We analyse integral field unit (IFU) data of ESO
148-IG002 to elucidate the relative fraction of the power sources (starburst,
AGN, shocks) and nature of shock activity in this galaxy.

\subsection{Observations and Data Reduction}
Our data for ESO 148-IG002 were taken with the Wide Field Spectrograph (WiFes) at
the Mount Stromlo and Siding Springs Observatory 2.3m telescope. WiFeS is a dual
beam, image slicing IFU described in detail by \cite{Dopita2007,Dopita2010}. Our
data consist of separate blue ($\sim$3500-5800 \AA) and red ($\sim$5500-7000
\AA) spectra with a resolution of R=3000 for the blue and R=7000 for the red.
This resolution corresponds to a velocity resolution of 100 km s$^{-1}$ at
H$\beta$ and 40 km s$^{-1}$ at H$\alpha$. Two pointings with WiFeS were taken on
July 28 and August 14 and 15 2009, with total exposure times of 3500s and 4000s
respectively. The region overlapping the two pointings would therefore have an
exposure time of 7500s. The seeing was on average 1.25" for our observations. The WiFeS field of view compared to ESO 148-IG002 is
shown in Figure \ref{aladin}. The IFU field consists of 25$\times$ 1" wide
slitlets, each of which is 38" long. The spatial pixel is 0.5" along the
slitlet axis and 1.0" in the spectral direction. Post reduction, data were
binned 2 pixels in the y-direction in order to increase the signal to noise and
produce a final resolution of 1"$\times$ 1".\\
The data were reduced and flux calibrated using the Flux Standard Feige 110 (a
white dwarf star) and the WiFeS pipeline. The WiFeS pipeline uses IRAF routines
adapted from the Gemini NIFS data reduction package and is briefly described in
\cite{Dopita2010}. Each observation was reduced into a data cube using the
process described in Rich et al. (2011). ESO 148-IG002 was observed using Nod
and Shuffle (N$\&$S) mode allowing sky subtraction to be performed using the
N$\&$S data. The sky was observed for the same amount of time as the object. The
data has bias frames subtracted and any residual bias level is accounted for
with a fit to unexposed regions of the detector. The mapping of the slitlets
required for spatial calibration is carried out by placing a thin wire in the
filter wheel whilst illuminating the slitlet array with a continuum lamp to
define the centre of each slitlet.
CuAr and NeAr arc lamps are used to wavelength calibrate the blue and red
spectra respectively.
Observations of a featureless white dwarf taken at similar air mass were used to
remove telluric absorption features from the red data cubes.

\subsection{Spectral Fitting}
To fit and remove stellar continuum from each spectrum, we used an automated
fitting routine, UHSPECFIT, which is described in \cite{Rupke2010b} and \cite{Rich2010}. The software was based on the fitting routines
of \cite{Moustakas2006}, which fits a linear combination of stellar
templates to a galaxy spectrum.  To fit our relatively high resolution data, we
used population synthesis models from \citet{GonzalezDelgado2005} as our stellar continuum
templates.
After the subtraction of the fitted continuum, lines in the resulting emission
spectra are fit simultaneously using a one- or two-component Gaussian, depending
on the goodness of the fit. The emission line fits are carried out in the same
manner as \citet{Rich2011}. All spectra were fit using both one- and two-component Gaussians. We adopt the fit with the lowest reduced $\chi^2$, however, properties of surrounding spaxels were also used to decide which fit to use.
In addition, all fits were checked by eye, to ensure they were reasonable.
We determined that $\sim$ 200 spaxels required a two-component Gaussian fit, compared to $\sim$ 140 spaxels whose emission lines were fit with a single component.
Each emission line component fit has a corresponding
redshift, flux peak and width. The instrumental resolution of 46 km s$^{-1}$ FWHM (full
width at half maximum) is subtracted in quadrature from the line fit FWHM to
compensate for instrumental broadening and converted to a $\sigma$ for the
purpose of our analysis.
The fitting routines used to fit both the continuum and emission lines made use
of the MPFIT package, which performs a least-squares analysis using the
Levenberg-Marquardt algorithm \citep{Markwardt2009}.
Example fits are provided in Figure 2, along with an example of a typical
spectrum across the entire spectral range.

\begin{figure}
\centering
\includegraphics[width=\linewidth]{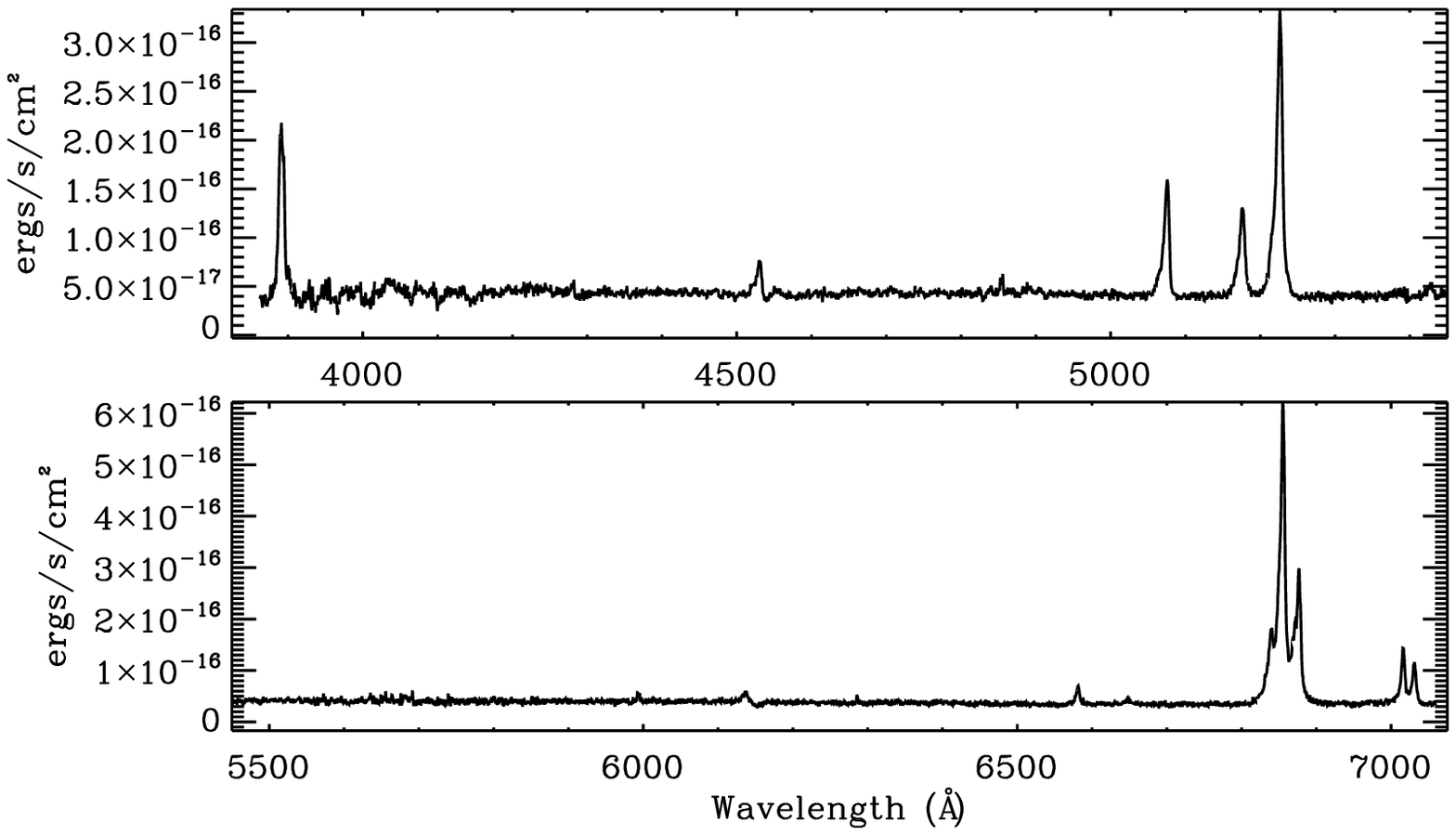}
\includegraphics[width=\linewidth]{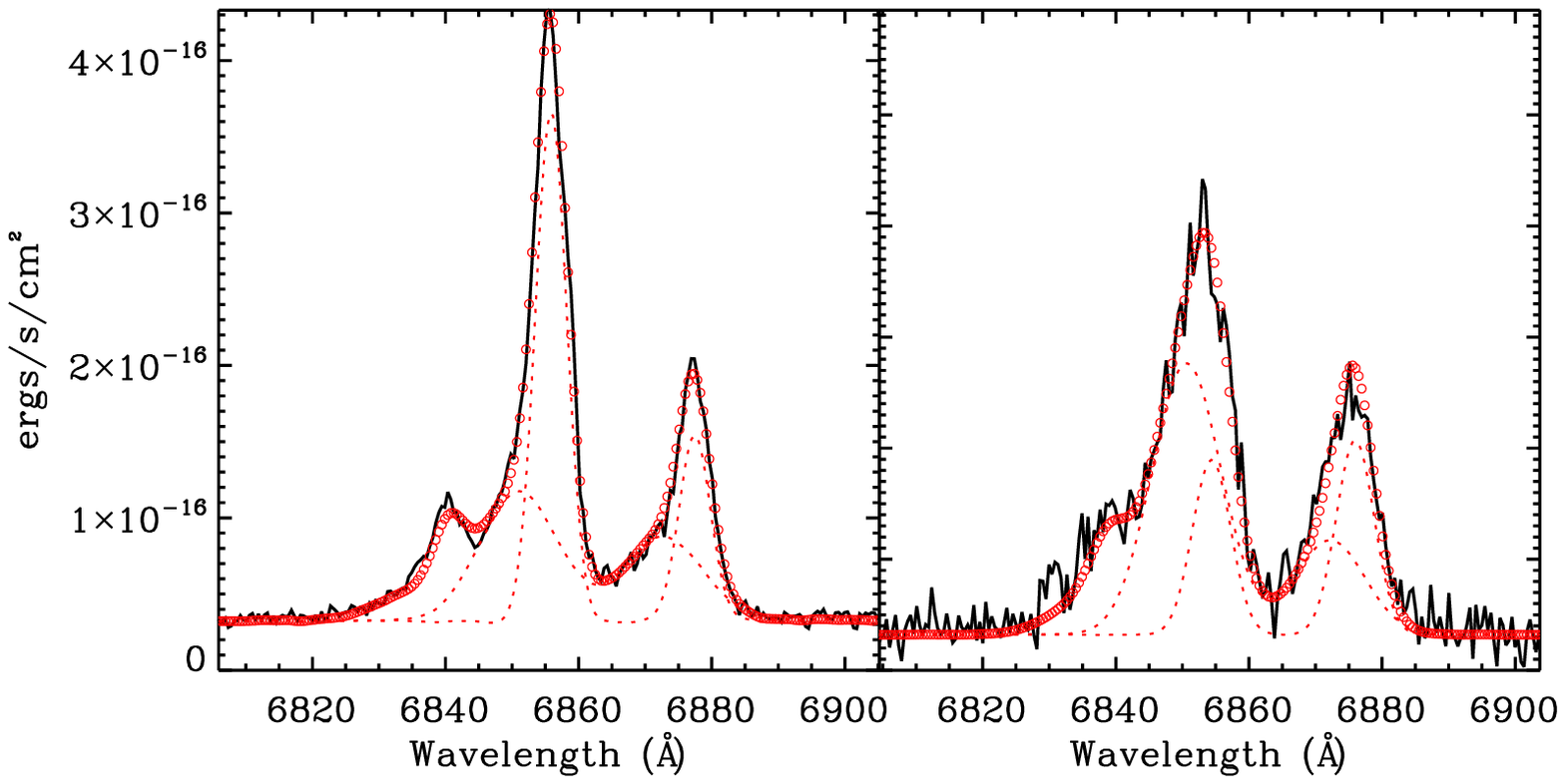}
\caption{The top two panels show an example of raw data for a single spectrum
across the whole spectral range. The bottom panel shows example
two-component fits for the H$\alpha$ and [NII] 6583 lines. The flux of the [NII]
6548 component is fixed to the [NII] 6583 flux (ratio 1:3; \citealt{Osterbrock2006}). We do not plot the [NII]6548 line components so the other fitted components can be seen clearly. The right hand spectrum is from the
northern nuclear region, whilst the left is from an off-nuclear region east of
the southern nucleus.  The dashed red lines show the two Gaussian functions which
together give the overall emission fit. The raw data are in black and the
continuum plus emission fit is in red. Both components are often well defined in one emission line but not in
another. Where the amplitude of an individual component is not more than five standard
deviations above the noise, it is not considered in subsequent
analysis.}\label{spectra}
\end{figure}

Errors in the parameters used in fitting Gaussian functions to the emission
lines are calculated by the MPFIT package, using the variance spectrum as input.
To analyse multiple component fits, we
minimise the $\chi^{2}$ for all lines simultaneously. As such all lines ([OII]3726, H$\beta$, [OIII]4959, [OIII]5007, [OI]6300, [NII]6548, H$\alpha$, [NII]6583, [SII] 6716. [SII]6731) have the same velocity and velocity dispersion. Assuming that all the gas
contained within a given spaxel is undergoing the same processes, we would
expect each emission line in the spectrum to have the same velocity dispersion.
By fitting all lines simultaneously, we decrease the overall uncertainty in the
velocity dispersion and maintain the ability utilise the multiple component
fits.

As our analysis of this kinematically complex system relies on a careful
decomposition of multiple velocity dispersions, it is crucial that more than one
Gaussian can be used in the emission line fit.

\section{Emission line gas properties}

\subsection{Line Ratio Maps}

\begin{figure*}
\centering
\mbox{\subfigure[HST]{\includegraphics[width=3.6cm,]{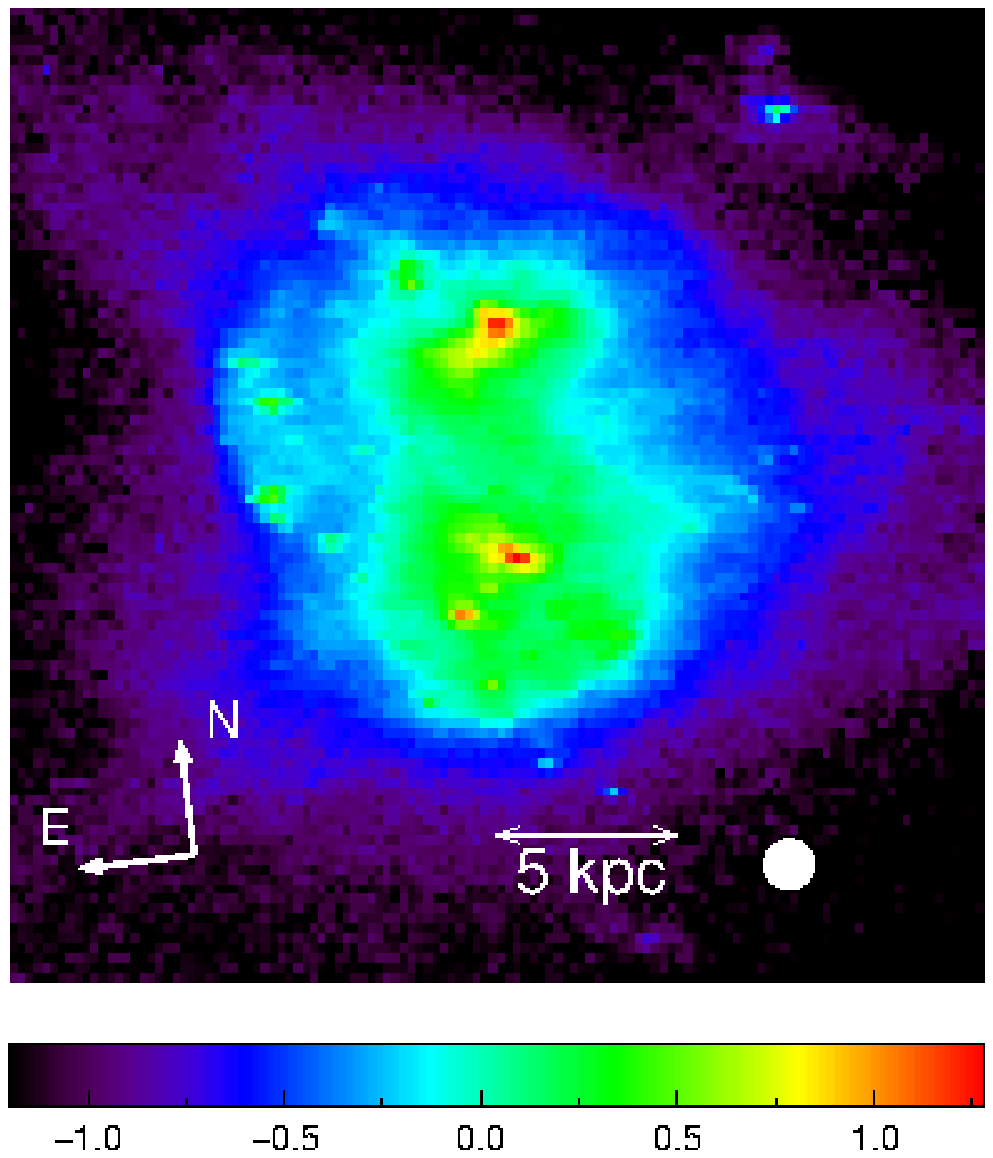}}\quad
\subfigure[H$\alpha$ ]{\includegraphics[width=3.6cm]{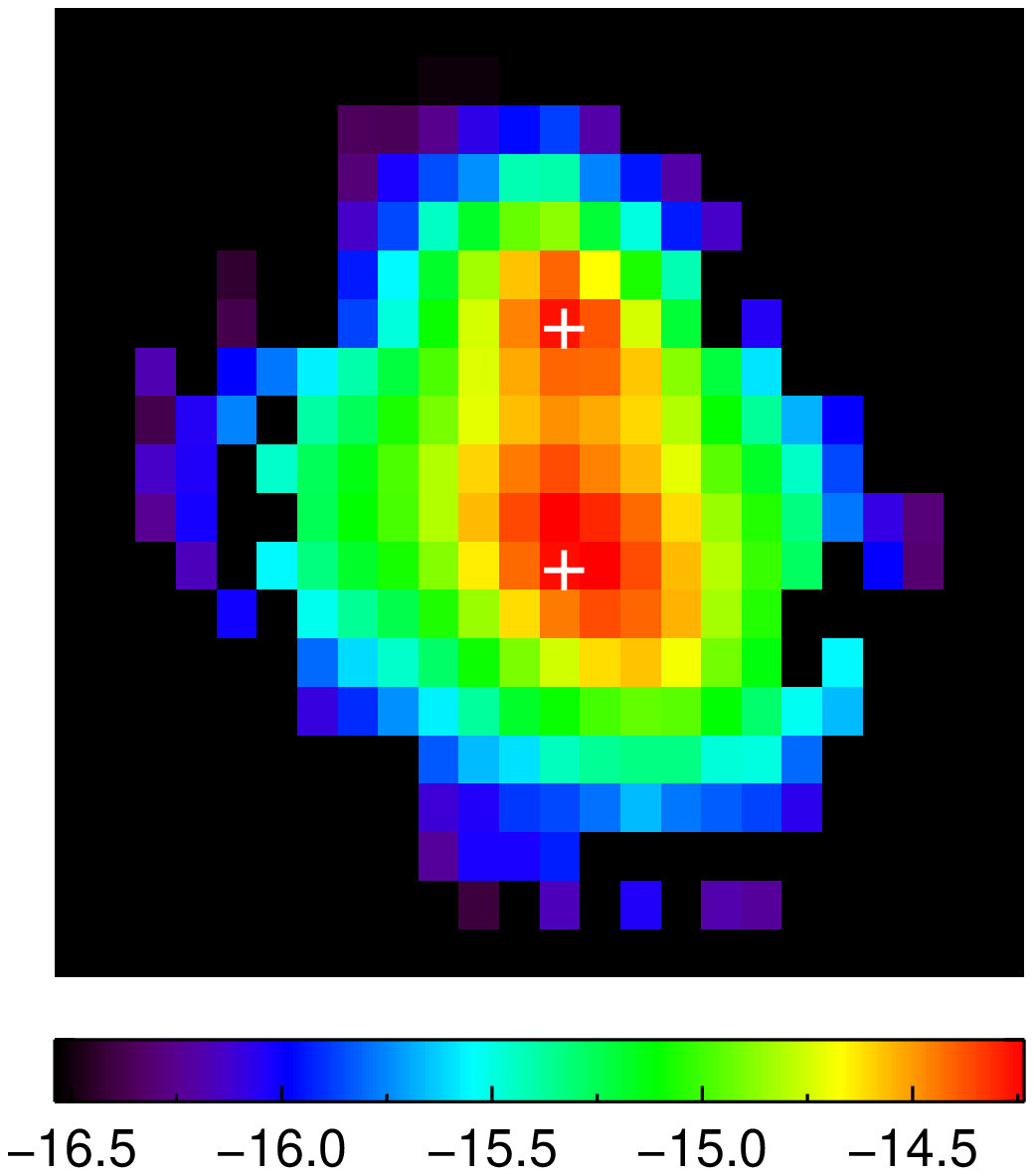}}\quad
\subfigure[\text{[NII]}]{\includegraphics[width=3.6cm]{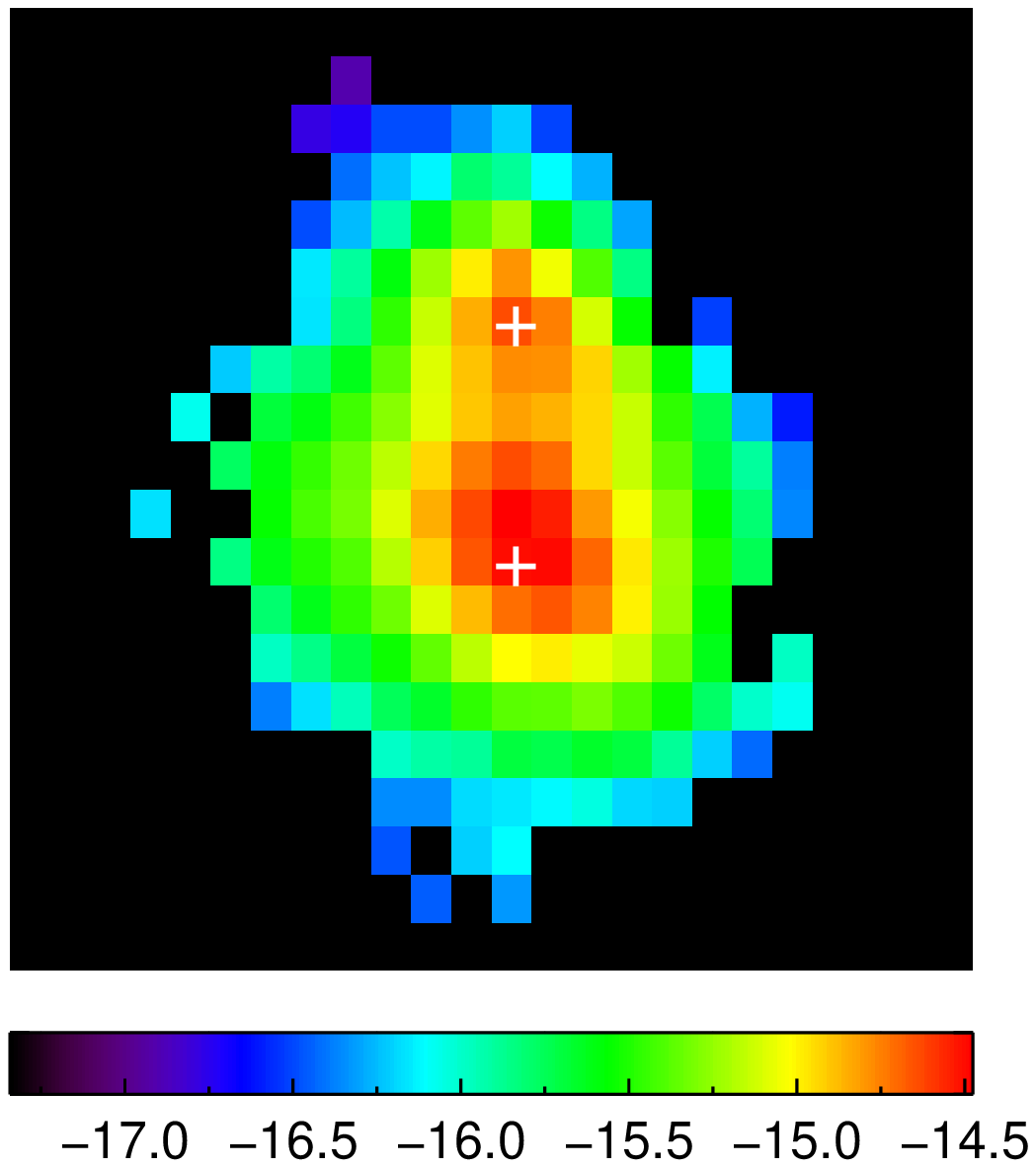}}\quad
\subfigure[\text{[NII]/H$\alpha$}]{\includegraphics[width=3.6cm]{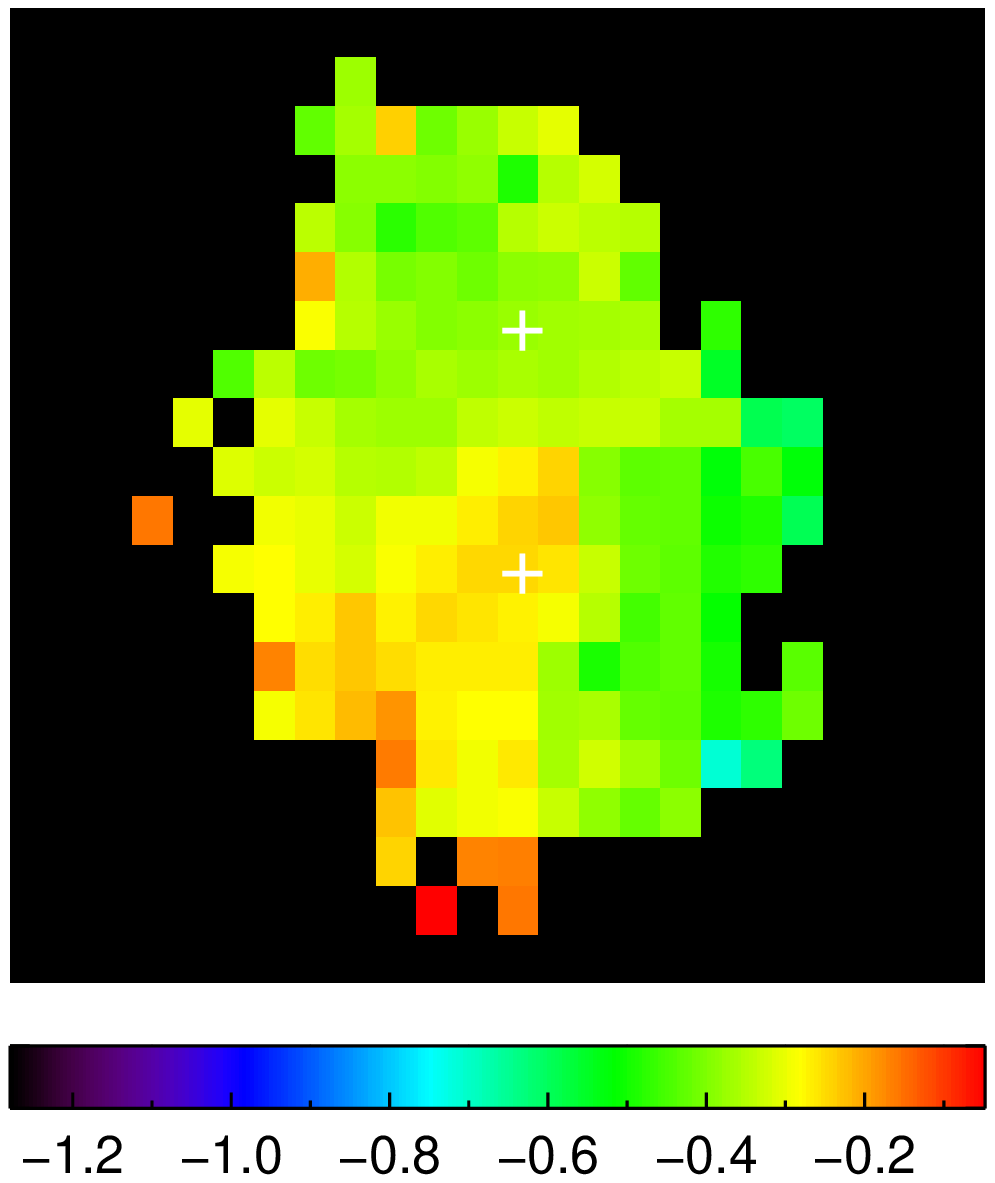}
}}
\mbox{\subfigure[HST]{\includegraphics[width=3.6cm,]{hst.eps}}\quad
\subfigure[H$\alpha$ error]{\includegraphics[width=3.6cm]{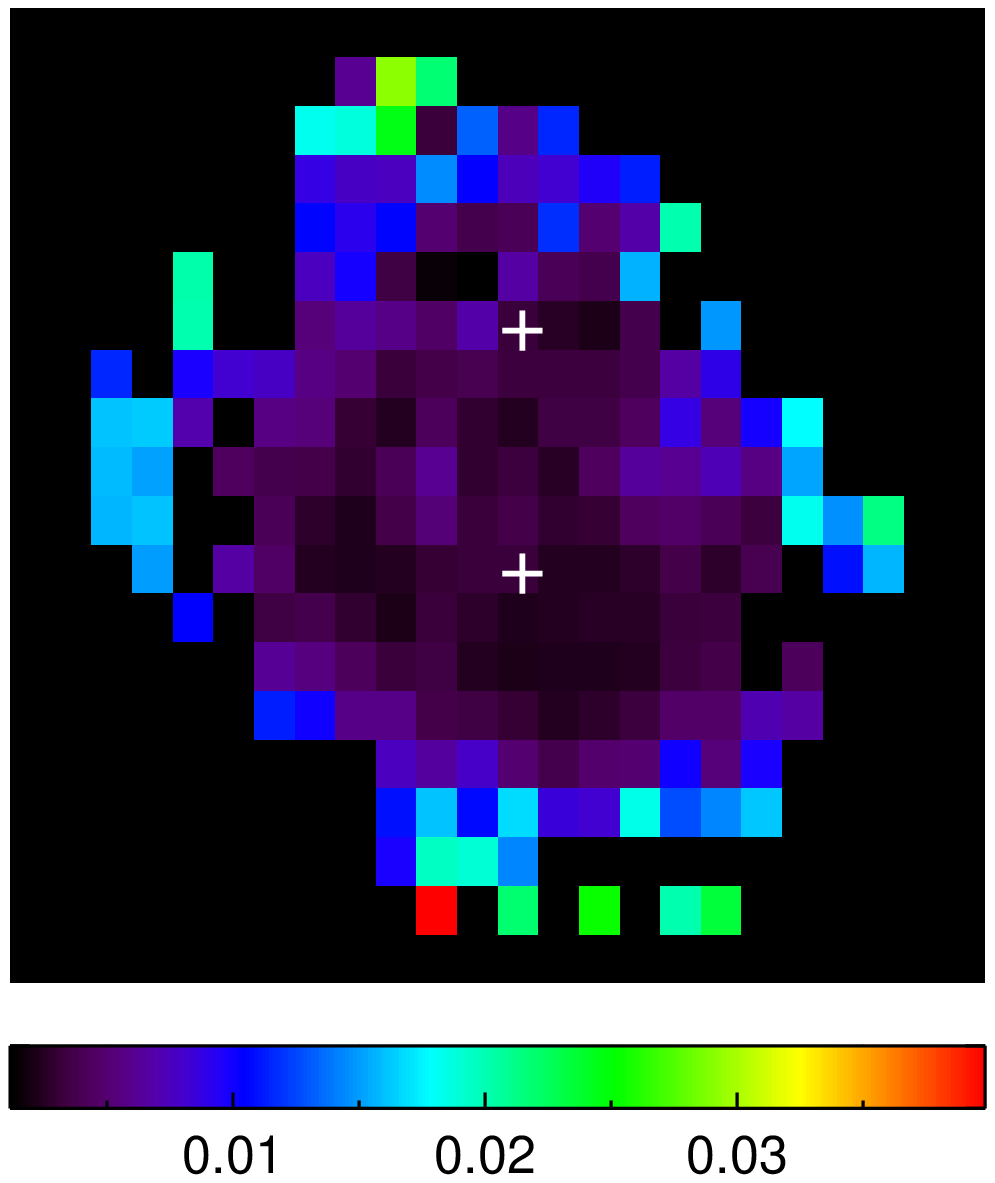}}\quad
\subfigure[\text{[NII]}
error]{\includegraphics[width=3.6cm]{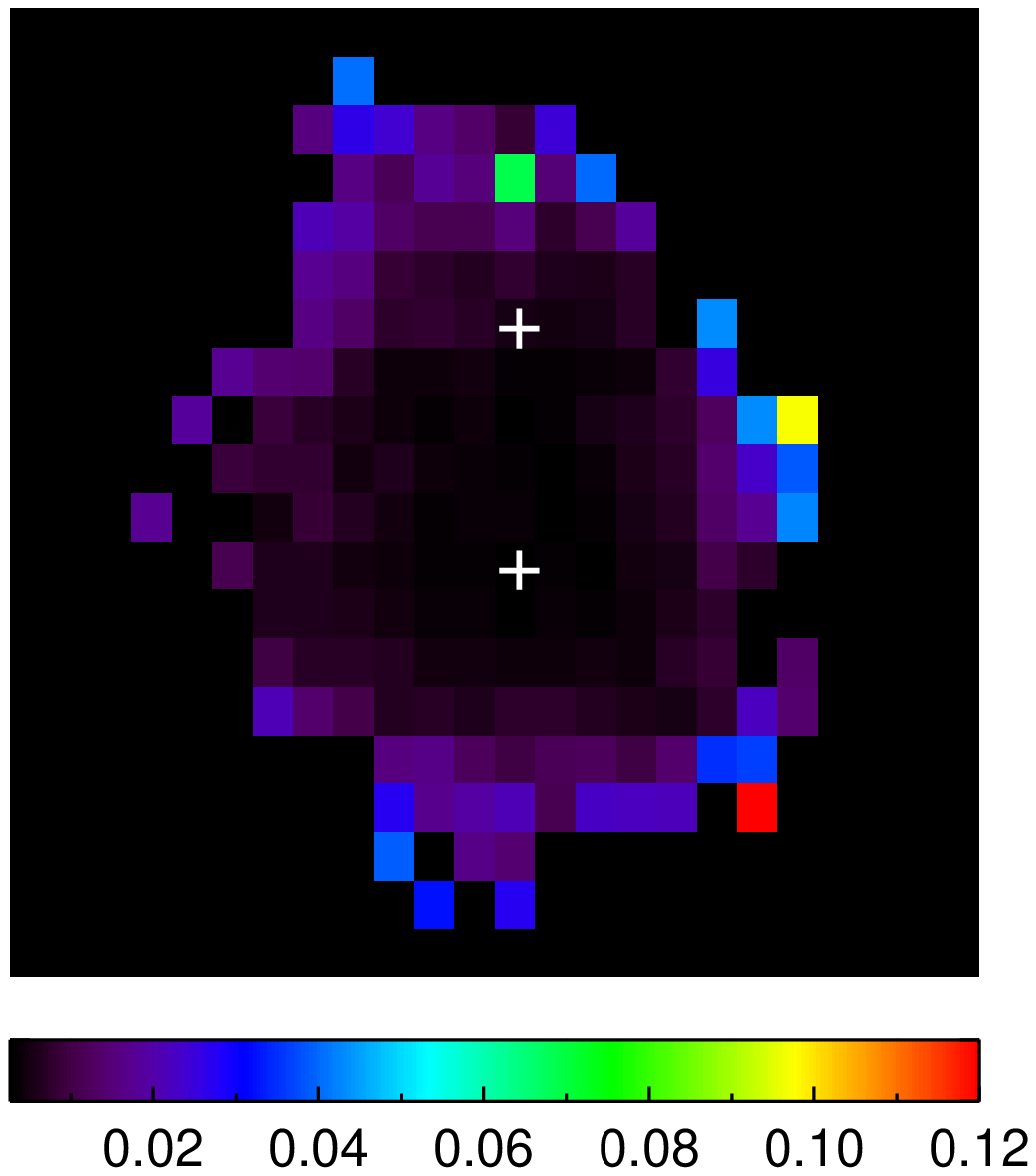}}\quad
\subfigure[\text{[NII]}/H$\alpha$
error]{\includegraphics[width=3.6cm]{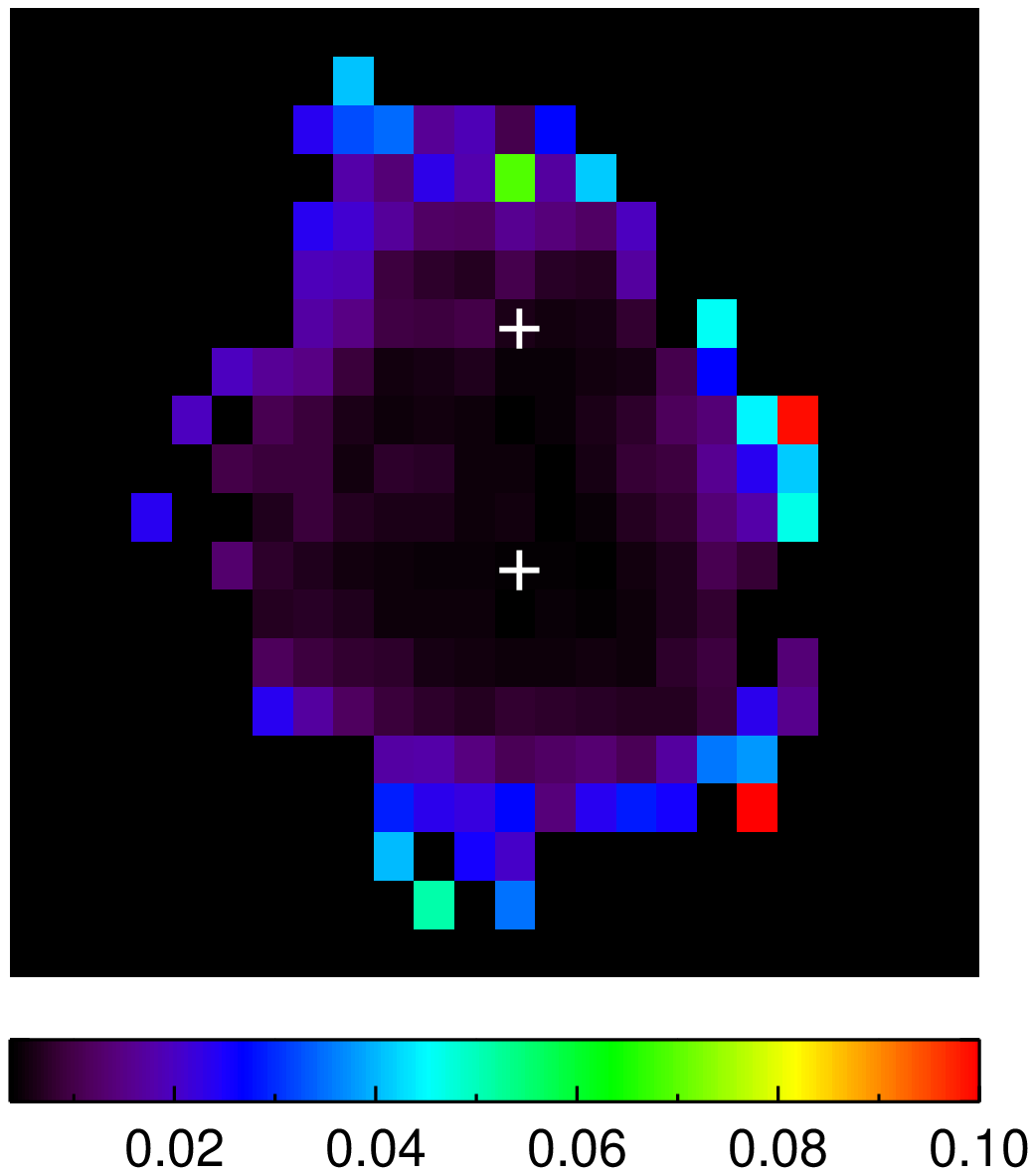}}}
\mbox{\subfigure[HST]{\includegraphics[width=3.6cm,]{hst.eps}}\quad
\subfigure[\text{[SII]}/H$\alpha$
]{\includegraphics[width=3.6cm]{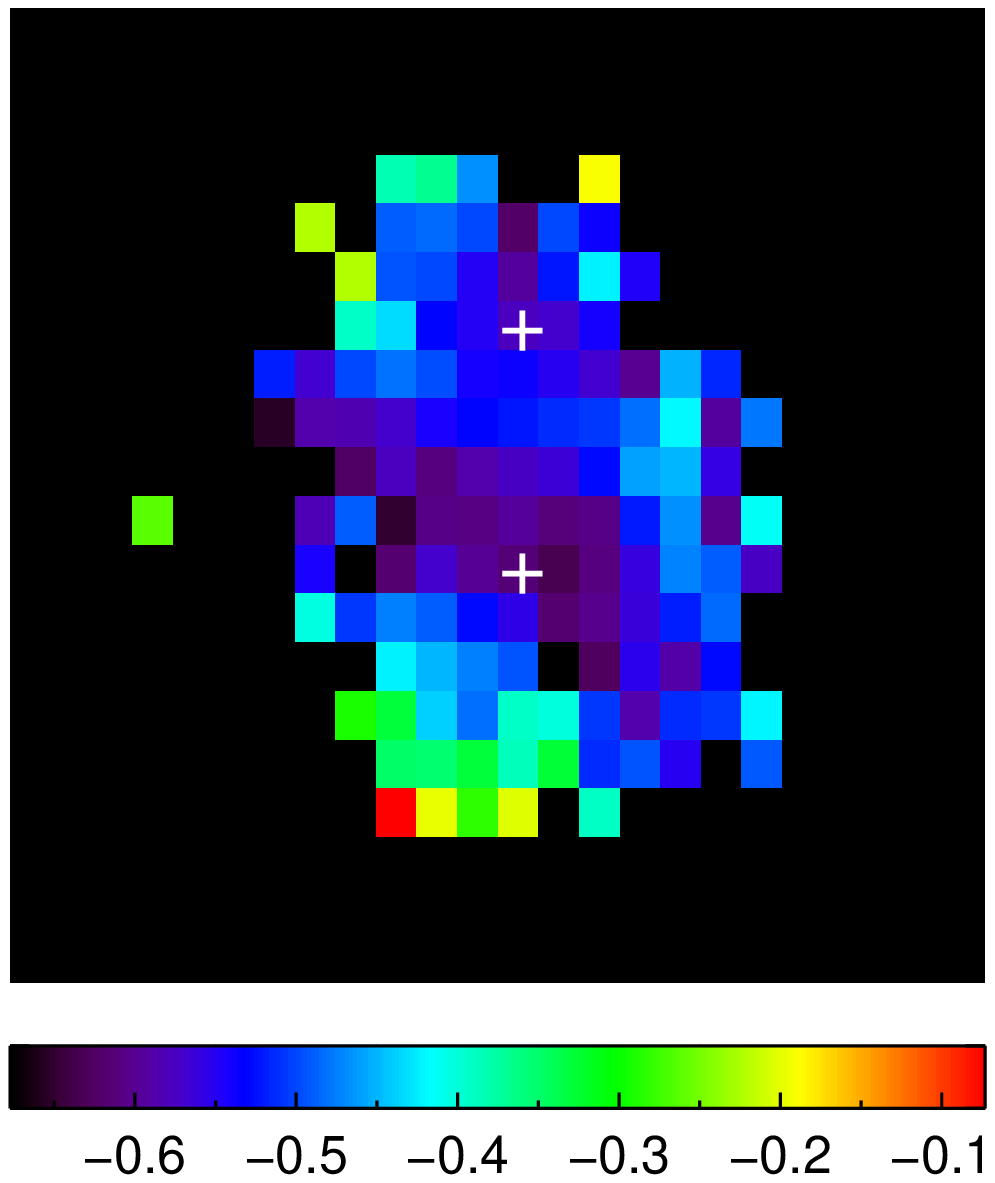}}\quad
\subfigure[\text{[OI]}/H$\alpha$]{\includegraphics[width=3.6cm]{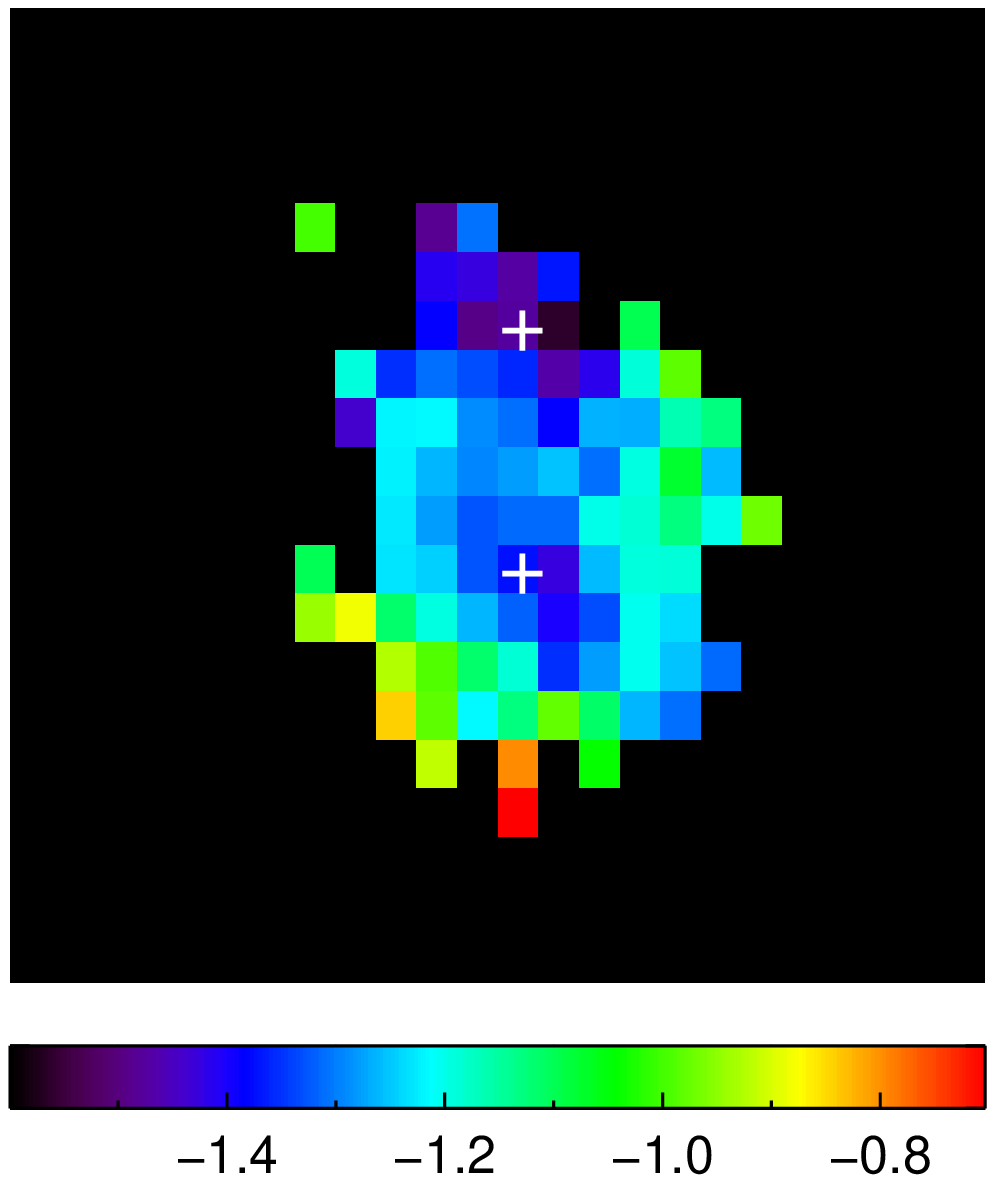}}
\quad
\subfigure[\text{[OIII]}/H$\beta$]{\includegraphics[width=3.6cm]{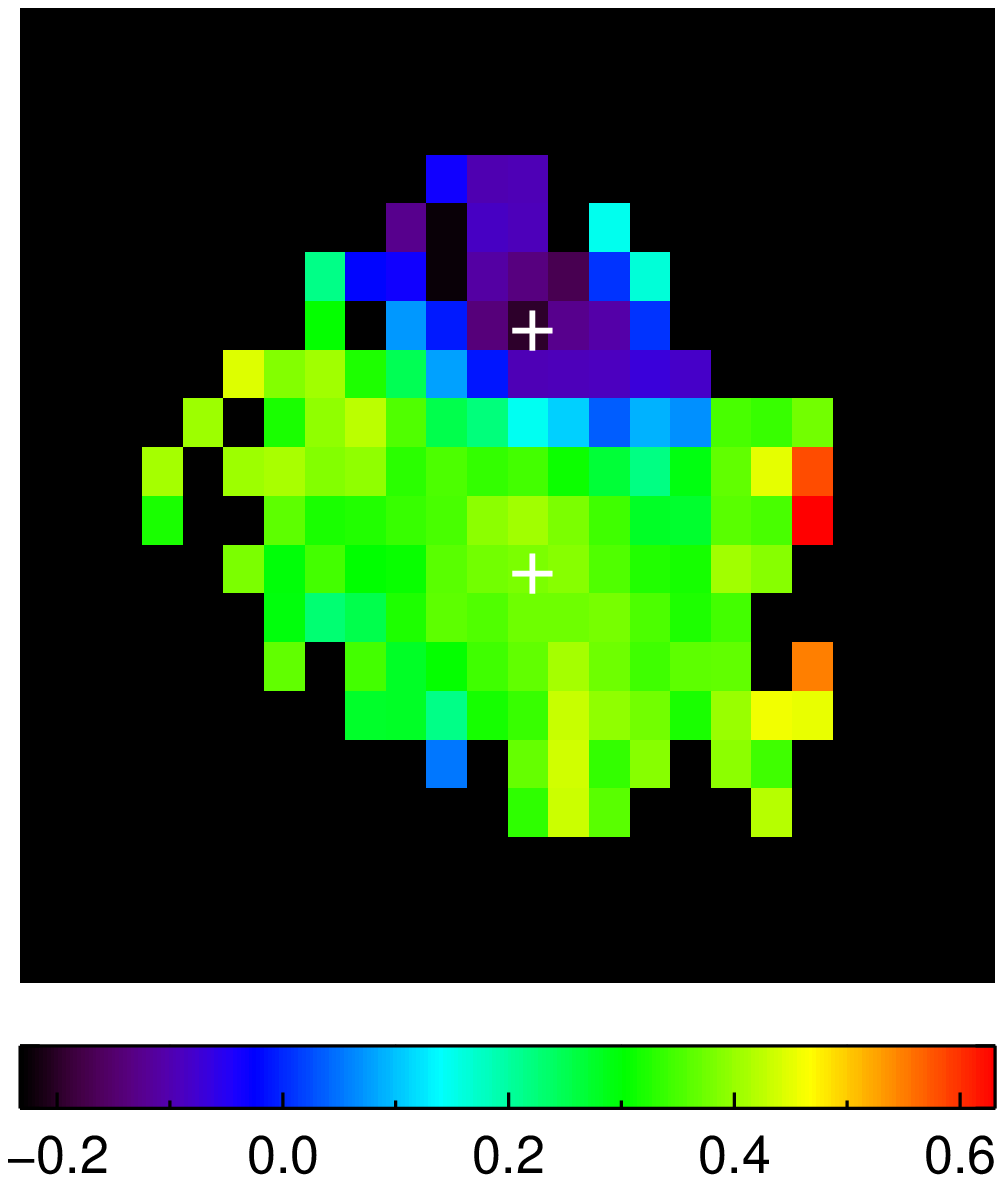}}}
\mbox{\subfigure[HST]{\includegraphics[width=3.6cm,]{hst.eps}}\quad
\subfigure[\text{[SII]}/H$\alpha$ error
]{\includegraphics[width=3.6cm]{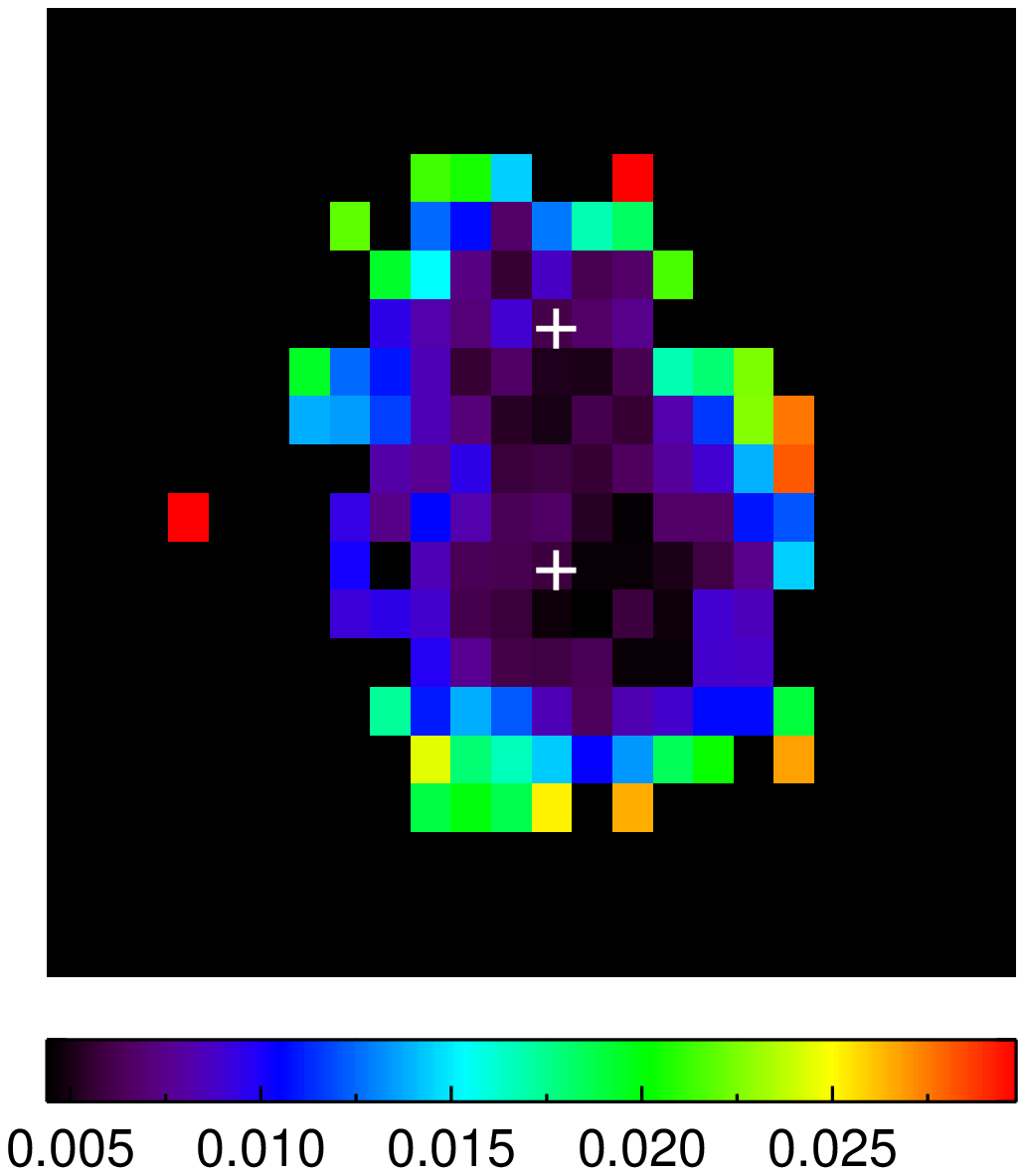}}\quad
\subfigure[\text{[OI]}/H$\alpha$
error]{\includegraphics[width=3.6cm]{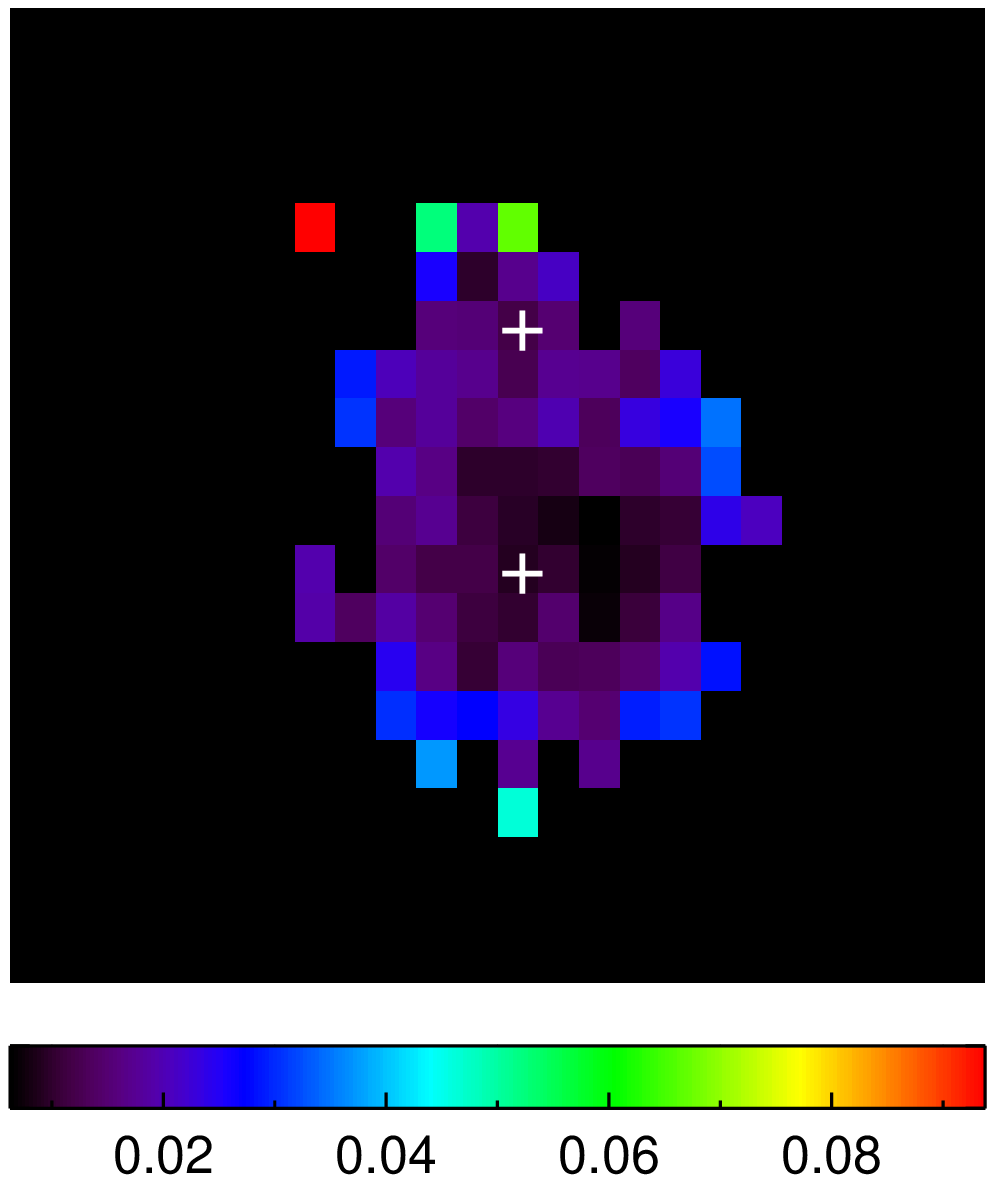}}\quad
\subfigure[\text{[OIII]}/H$\beta$
error]{\includegraphics[width=3.6cm]{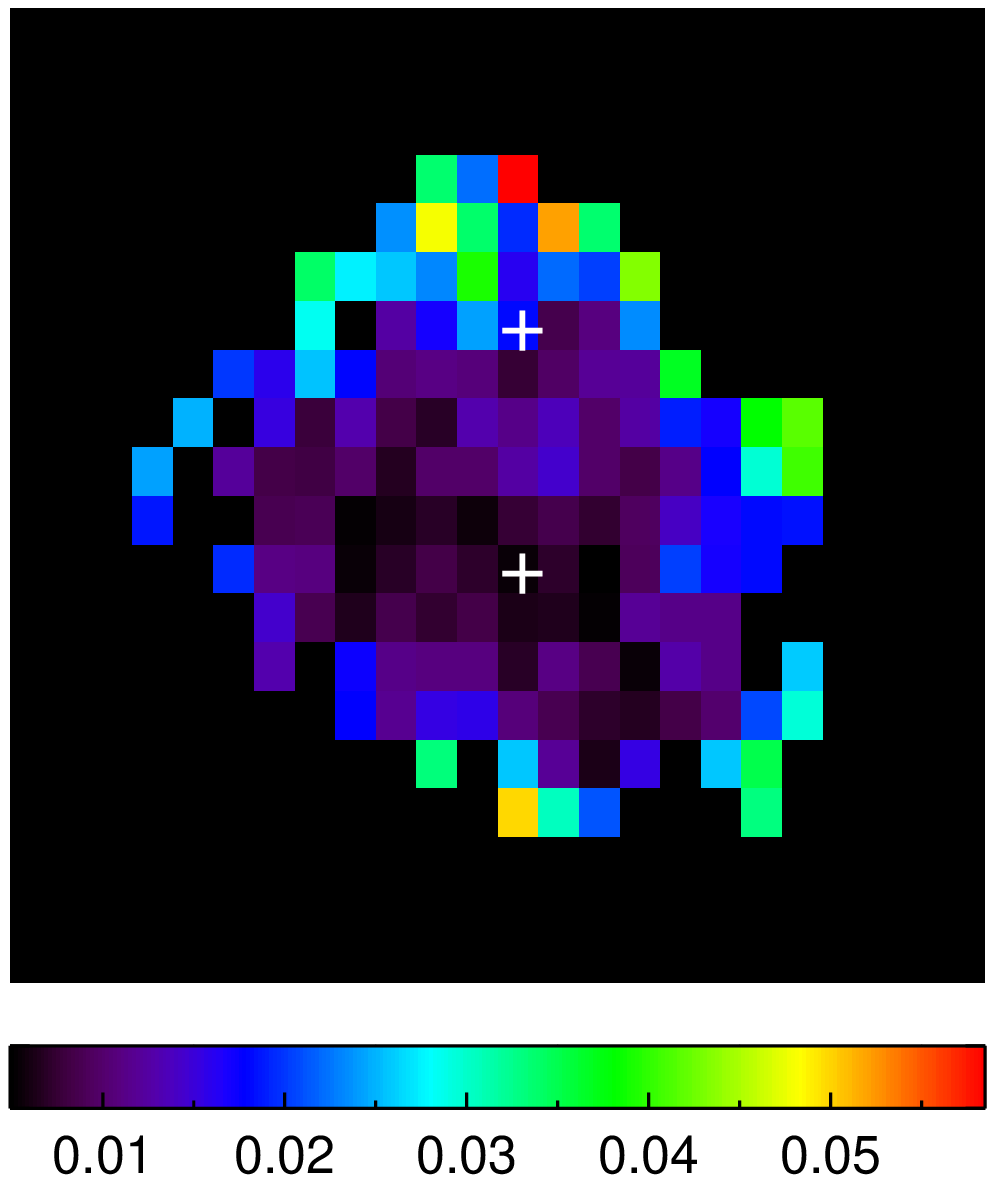}}}
\centering
\caption{Emission line ratio maps using total fluxes (the sum of both Gaussians if a double fit is made).
All scale bars show the log of the value
indicated by the captions. The top images show the log of the total emission of
[NII] and H$\alpha$ (in units of ergs s$^{-1}$ cm$^{-2}$ \AA$^{-1}$) of ESO
148-IG002, and the ratio of the two. In both the H$\alpha$ and [NII] maps, two
peaks in the emission can be seen. This most likely corresponds to the two
nuclei from the original galaxies (as determined by the peaks in the HST image),
shown by white crosses on the maps. Underneath the images their corresponding
relative error maps are also shown on a logarithmic scale. Images (j), (k), and
(l) are the line ratio maps corresponding to the diagnostic diagrams for
log([SII]/H$\alpha$), log([OI]/H$\alpha$) and log([OIII]/H$\beta$). The
left most panels are the HST image of ESO 148-IG002 shown in Figure 1, with the
same dimensions and orientation as the other maps, for reference. The white
circle on the bottom right corner of the HST image is an estimate of the seeing
for our WiFeS observations.} \label{ratiomaps}
\end{figure*}

We examine the maps of the ratios of the total flux of [NII]$\lambda$6583,
[SII]$\lambda\lambda$6716,6731 and [OI]$\lambda$6300 to H$\alpha$ in Figure
\ref{ratiomaps} as a first step towards understanding the processes at work in
ESO 148-IG002. Where two Gaussians are used to fit an individual emission line, the total flux is taken to be the
sum of the two component fluxes. These ratios are sensitive to metallicity as well as ionization
parameter and have the advantage that they do not require reddening corrections.
Line ratio values corresponding to each spaxel are also plotted on standard
diagnostic diagrams in Figure \ref{BPT}. In general, the weaker the line ratio,
the greater the contribution of pure star formation to the ionizing spectrum
\citep{Kewley2006}. In Figure \ref{ratiomaps} we also show the ratio of
[OIII]$\lambda$5007 to H$\beta$, used in all diagnostic diagrams of Figure
\ref{BPT}. We represent the errors in the line ratio values by mapping the
uncertainty in each spaxel on a logarithmic scale.

The [NII]/H$\alpha$ ratio is sensitive to metallicity
\citep{Kewley2002,Denicolo2002,Pettini2004,Kewley2008}. The higher
[NII]/H$\alpha$ ratio in the southern corner of the galaxy could thus correspond
to a higher metallicity clump, however the AGN in the southern nucleus (bottom
white cross in Figure 3) may also contribute to the [NII]/H$\alpha$ ratio. The
other two ratios, [SII]/H$\alpha$ and [OI]/H$\alpha$, are more sensitive
to a hard radiation field from AGN or shocks. Higher ratios in the outer
regions, such as in these two maps, usually correspond to shocked regions (e.g.
\citealt{Rich2010,Monreal-Ibero2010,Sharp2010,Rich2011}. However, in this case there are only
$\sim$2 spaxels detected in the outer regions with elevated [SII]/H$\alpha$ and
[OI]/H$\alpha$ ratios, so we cannot conclude whether shocks are affecting the area.

\subsection{Diagnostic Diagrams}
To better understand the contributing power sources in this galaxy, we use line
ratio diagnostic diagrams. Diagnostics using the [NII]/H$\alpha$,
[SII]/H$\alpha$ and [OI]/H$\alpha$ against [OIII]/H$\beta$ ratios were first
employed by \cite{Baldwin1981} and \cite{Veilleux1987} to
distinguish the likely ionizing source of emission line gas in galaxies. \cite{Kewley2001} used a combination of stellar population synthesis models and
self-consistent photoionization models to determine a theoretical ``maximum
starburst line'' on the diagrams which indicates the theoretical upper limit
given by pure stellar photoionization models. The diagrams have been
subsequently updated by \cite{Kauffmann2003} and \cite{Kewley2006} to
include empirical lines dividing pure star-forming from Seyfert-HII composite
galaxies and Seyferts from LINER galaxies respectively.

\begin{figure*}
\centering
\includegraphics[width=0.9\linewidth]{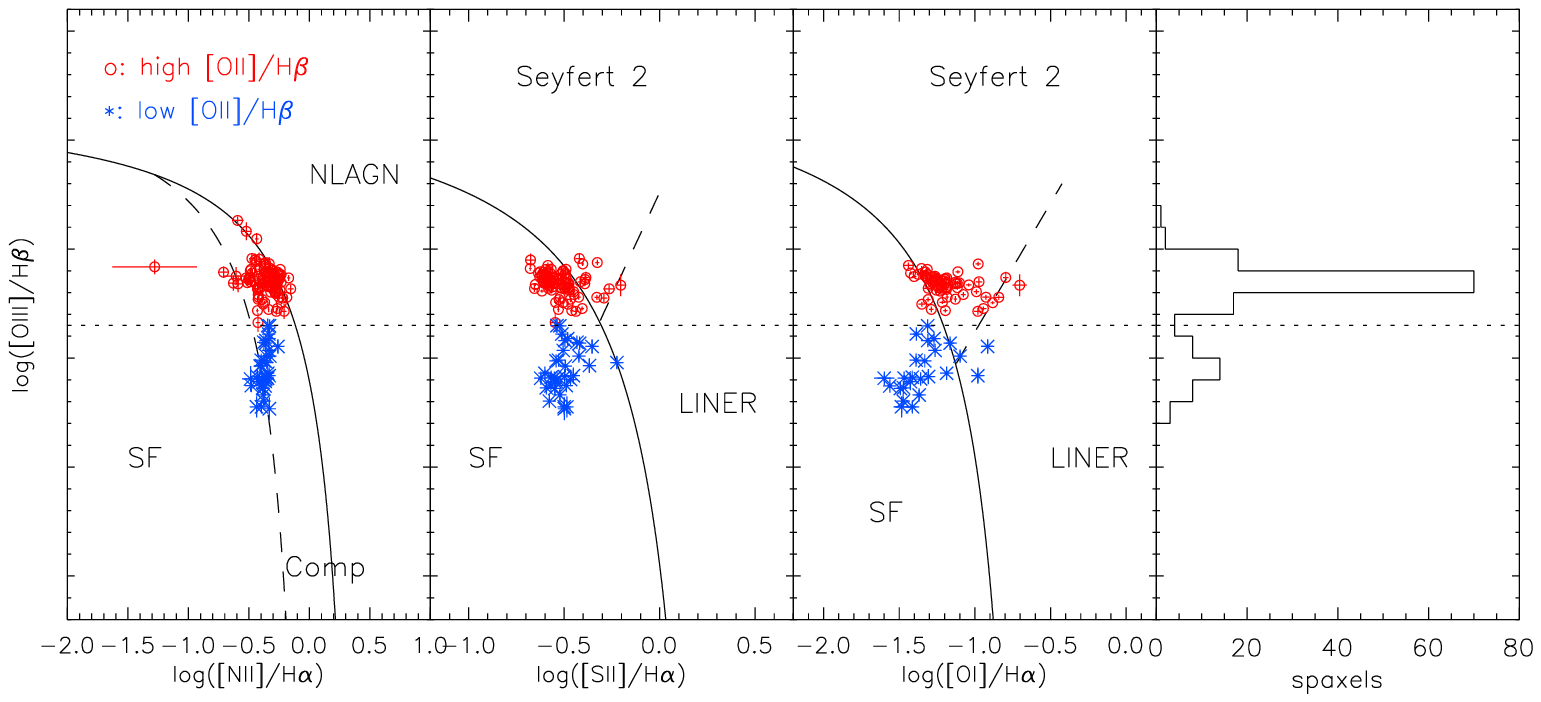}
\mbox{\subfigure[HST]{\epsfig{figure=hst.eps,width=4.5cm,}}\quad
\subfigure[Location of spaxels]{\epsfig{figure=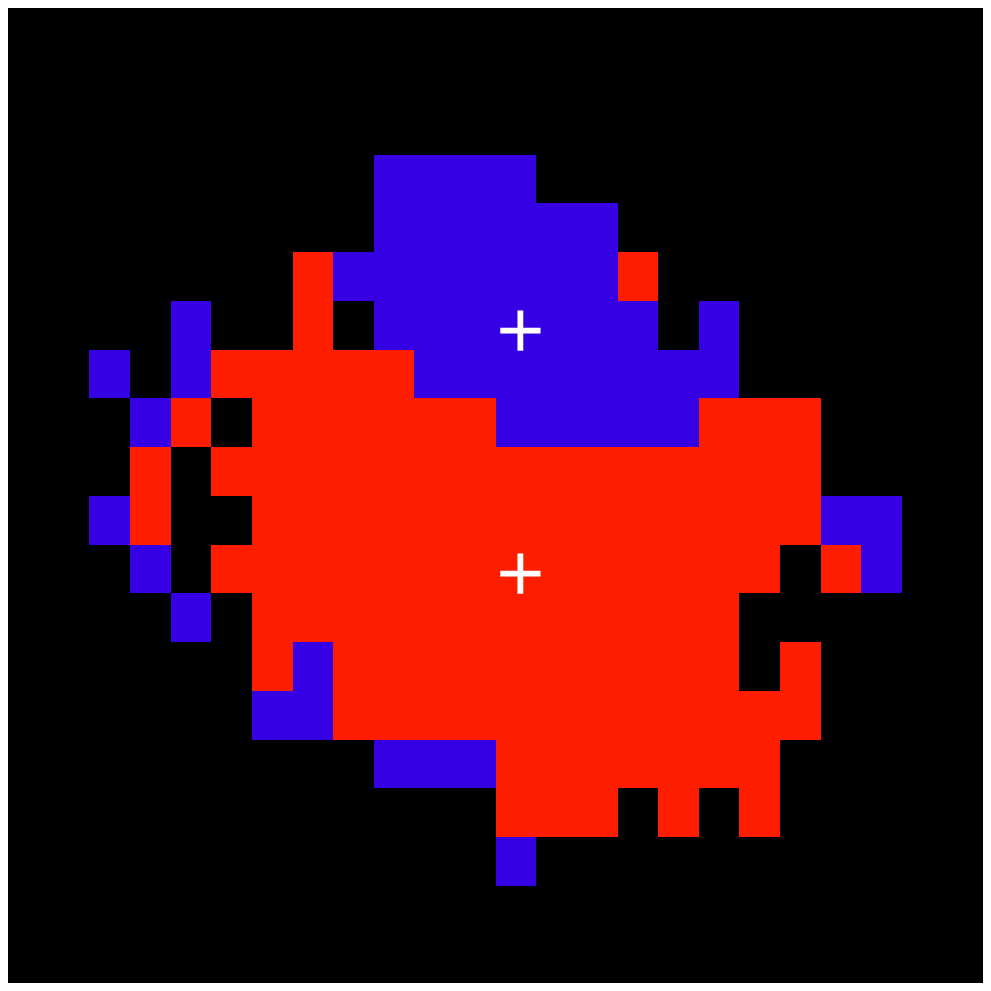,width=4.5cm}}}
\caption{Emission line ratio diagnostic diagrams of each individual fitted
emission line. Here total fluxes for each line are used (as in Figure 3). Black solid curves form an upper limit for star-forming
galaxies as derived by Kewley et al. (2001). The dashed line on the [NII]
diagram is the empirical Kauffmann et al. (2003) boundary below which galaxies
are classified as star-forming. The dashed lines on the [SII] and [OI] diagrams
were derived by Kewley et al. (2006) to empirically separate Seyfert 2 galaxies
and LINERs. In the leftmost panel, NLAGN represents narrow emission-line AGN (Seyfert 2 plus LINERs); Comp
represents starburst-AGN composites. We apply a minimum S/N cut of 5 in every diagnostic line. Typical errors are 0.009 dex for log([NII]/H$\alpha$) and log([SII]/H$\alpha$), 0.016 dex for log([OI]/H$\alpha$) and 0.012 for log([OIII]/H$\beta$). The
spaxels in each diagram reflect a mix of ionization sources. The spaxels are distributed in two clumps in [OIII]/H$\beta$ as shown in the histogram of the top right panel. The division between the two clumps is shown across the top panel by the dotted line, marking the minimum in the bimodal [OIII]/H$\beta$ distribution. The points are also colour coded based on this division.
The spatial position of the clumps are shown in the bottom right panel,
with the red and blue colors matching those from the BPT diagram. This could indicate that gas from the northern nucleus (blue)
has a higher metallicity.}\label{BPT}
\end{figure*}

We show the  diagnostic diagrams for ESO 148-IG002 in Figure \ref{BPT}. We include the emission line fluxes from each spaxel with $S/N>5$  in all relevant lines. This
procedure allows us to classify the dominant energy source in each spaxel. In
all three diagrams the lower left hand section of the plot traces
photoionization by HII regions. The solid curve traces the upper theoretical
limit to the pure HII region contribution measured by \cite{Kewley2001}. The
observed dashed line in the [NII] diagnostic provides an empirical upper limit
to the pure HII region sequence of Sloan Digital Sky Survey galaxies measured by
\cite{Kauffmann2003}. The region lying between these two lines represents
objects with a composite spectrum which is a mix of HII region emission and a
stronger ionizing source. LINER-like emission lies to the right-hand side of the
diagrams whilst contribution from a Seyfert AGN will push the spaxels upwards on
all three diagrams. If the galaxy is influenced by shocks, the line ratios can
be moved towards the LINER region \citep{Rich2010,Rich2011}. From the [NII]
diagnostic plot it is clear that there are very few purely star forming regions
in this galaxy.

The large number of SDSS galaxies ($\sim$45,000), allowed \cite{Kewley2006} to separate two clear branches on the [SII]/H$\alpha$ and
[OI]/H$\alpha$ diagnostic diagrams, empirically deriving a boundary between
Seyfert 2s and LINERs seen as the dashed lines on these diagrams.
LINERs have a harder ionizing radiation field and lower ionization parameter
than Seyfert galaxies, making the [SII]/H$\alpha$ and [OI]/H$\alpha$ diagrams
ideal for separating Seyferts and LINERS. The [SII] and [OI] emission lines are
produced in the partially ionized zone at the edge of the nebula. For hard
radiation fields, this zone is large and extended.
The [NII]/H$\alpha$ ratio only weakly depends on the hardness of the radiation
field, is much more dependent on the metallicity of the nebular gas and thus
cannot be used to separate Seyfert and LINER galaxies. The [SII] and [OI]
diagnostic diagrams in Figure \ref{BPT} indicate that the composite HII regions
seen in the [NII] diagram could be influenced by AGN activity as the spaxels are
spread upwards towards the Seyfert 2 branch.

ESO 148-IG002 has a bimodal distribution of [OIII]/H$\beta$ line ratios, which we have visually separated in the top panel of Figure \ref{BPT}.
The location of the spaxels with high and low [OIII]/H$\beta$ ratios on the BPT diagram, combined with their physical location in the galaxy, could suggest
that the northern nucleus has a higher metallicity than the southern nucleus.

\section{Kinematic Properties}
The power source of the emission line ratios is unclear from the
diagnostic diagrams alone. In this section we try to better understand the
dynamics as well as the different power sources of ESO 148-IG002. Here we make
most use of the two Gaussian component fits. As all emission lines in a spectrum
were fit simultaneously with one or two Gaussian components, all lines per spaxel have the same
velocities, and velocity dispersions, which are used in this section. In data
with high spectral resolution, one can analyse the distribution of velocity
dispersions to separate the galaxy's ionizing sources.
\subsection{Velocity Dispersion}
\begin{figure}
\centerline{\includegraphics[width=0.9\linewidth]{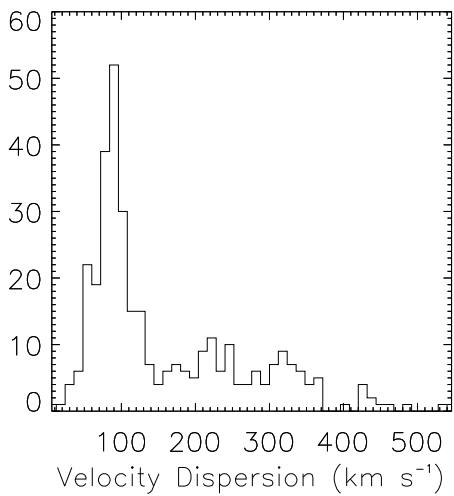}}
\caption{Velocity dispersion distribution from every individual Gaussian
component fit. The distribution has a peak around $\sim$ 100 km s$^{-1}$. Pure star-forming regions have velocity dispersions of $\sim 40$ km s$^{-1}$ \citep{Epinat2010}. Therefore it is likely that spaxels with velocity dispersions less than the local minimum at $\sigma<155$ km s$^{-1}$ are likely to be dominated by star formation, but also have contribution from shock or AGN. The extended distribution out to higher velocity dispersions could represent an increasing contribution from AGN activity.}\label{hist}
\end{figure}

Velocity dispersion, $\sigma$, is the result of the superposition of many line
profiles, each of which has been Doppler shifted and broadened because of the
gas motions within the galaxy. The velocity dispersions used for our
analysis have had the instrumental resolution ($\sim$ 40 km s$^{-1}$ FWHM in the red)
subtracted in quadrature to account for instrumental broadening.
In a complex galaxy such as ESO 148-IG 002, a large proportion of the gas may be
affected by more than one source of ionization. In this section we focus on the
velocity dispersion of individual emission line components, determined from the
two component fits, as a useful way to probe the processes influencing each
spaxel. HII regions correspond to a low velocity dispersion, typically of a few
tens of km s$^{-1}$ \citep{Epinat2010}, while slow shocks associated with
galactic winds have velocity dispersions $\ga 100$ km
s$^{-1}$\citep{Rich2010,Rich2011}, and the presence of an AGN can produce even
broader lines \citep{Wilson1985}. In Figure \ref{hist} we show the velocity dispersion
distribution from every component fit by our routine with S/N$>$5. The spectrum from every
spaxel was fit with either one or two Gaussian curves with each emission line in a spectrum
having the same fixed dispersion values (one or two depending how many components were needed), and
all components (for every spaxel) are shown.

\begin{figure*}
\centering
\mbox{\subfigure[HST]{\epsfig{figure=hst.eps,width=4cm,}}\quad
\subfigure[$\sigma<155$ km
s$^{-1}$]{\epsfig{figure=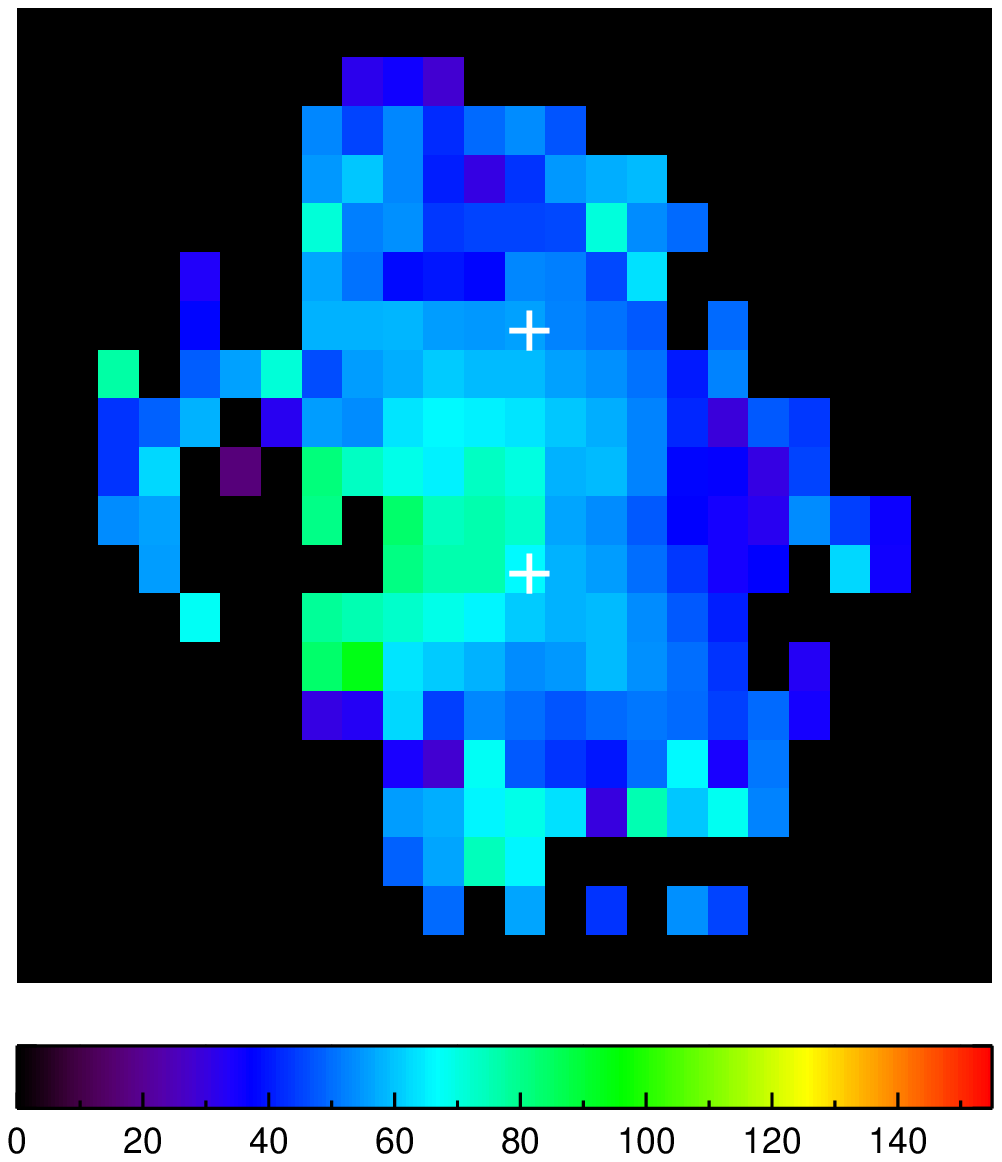,width=4cm}}\quad
\subfigure[$\sigma>155$ km s$^{-1}$]{\epsfig{figure=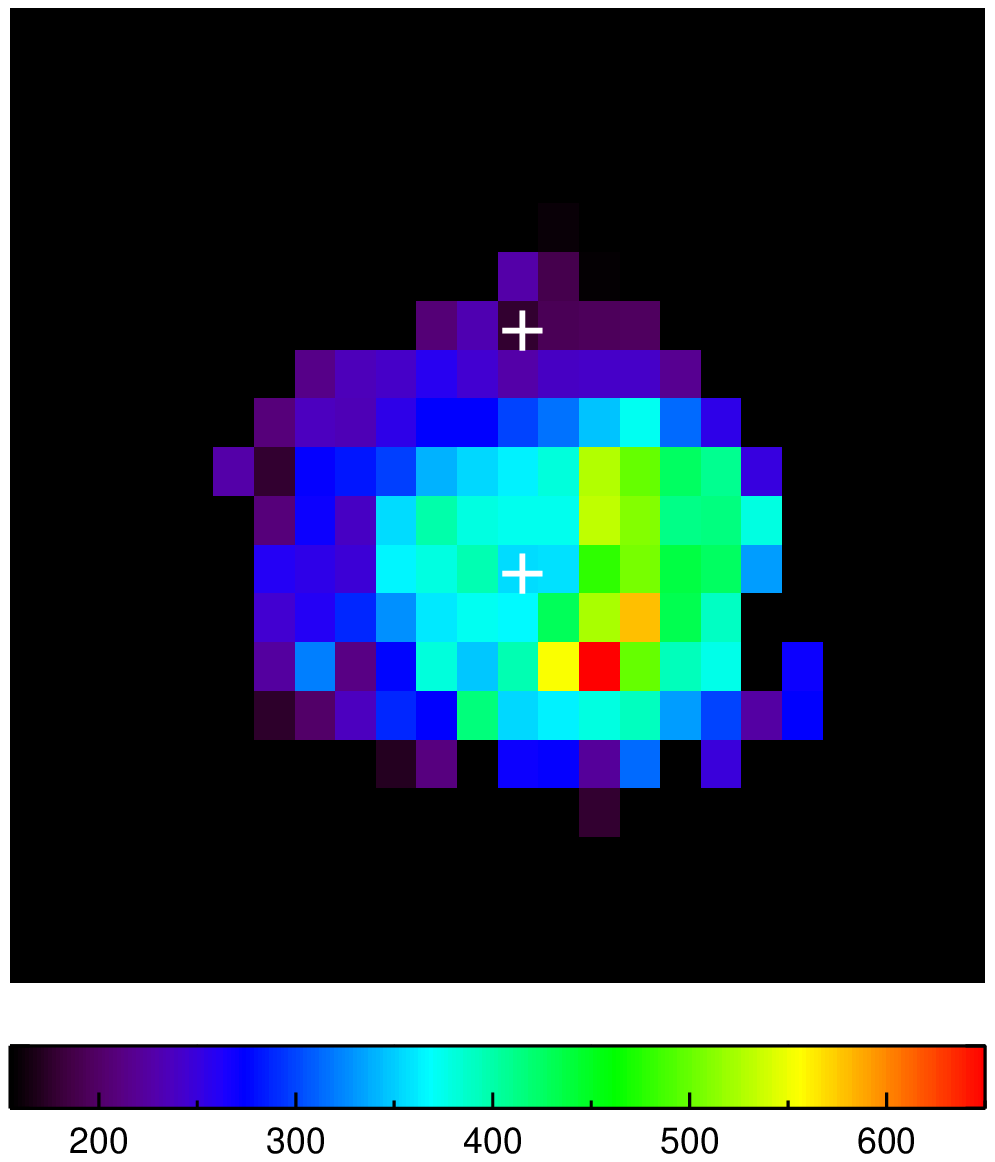,width=4cm}}}
\caption{Velocity dispersions in km s$^{-1}$ of the two components, with grouping
determined by the cut off. Left to right corresponds to increasing velocity
dispersion. For spatial comparison the HST image has been aligned and placed in
the leftmost panel. Where a spaxel contains two components fit which fall in the same velocity dispersion bin, we show a flux weighted average.
Spaxels which do not appear in the $\sigma > 155$ km s$^{-1}$ figure are the cases where either only one component is fit, and that component is narrow,
or two components are fit and both are narrow. The low velocity dispersion gas is extended over the whole galaxy, whereas the high velocity dispersion
gas is only found in the southern region. }\label{vmaps}
\end{figure*}

We establish a cutoff of $\sigma=155$ km s$^{-1}$ between the low-$\sigma$ HII
region emission and higher $\sigma$ emission by shocks and/or AGN, as there is a local minimum in the velocity dispersion distribution at this value (Figure \ref{hist}). The choice of
appropriate cutoffs was also influenced by the velocity dispersion maps, shown
in Figure \ref{vmaps}. The fact that the different components occupy different
regions indicates that they may be influenced by different phenomena. In some
cases a spaxel has more than one component in the same velocity dispersion bin.
In this case, to create the velocity dispersion maps, the $\sigma$ values were
averaged (weighted by the component's H$\alpha$ flux) over the two components within the same bin. The high velocity
dispersion gas is located to the South of the system, where the AGN lies.
However, the peak velocity dispersion is offset from the nucleus. This is
consistent with the idea that the high velocity dispersion gas is powered in
part by an AGN.  As shocks tend to correspond to a larger velocity dispersion
than that seen on the outskirts of this galaxy, it is unlikely that shocks are
contributing significantly to the emission spectrum.

\subsection{Velocity Maps}
\begin{figure*}
\centering
\mbox{\subfigure[HST]{\epsfig{figure=hst.eps,width=4cm}}\quad
\subfigure[Velocity(km s$^{-1}$): $\sigma<155$ km s$
^{-1}$]{\epsfig{figure=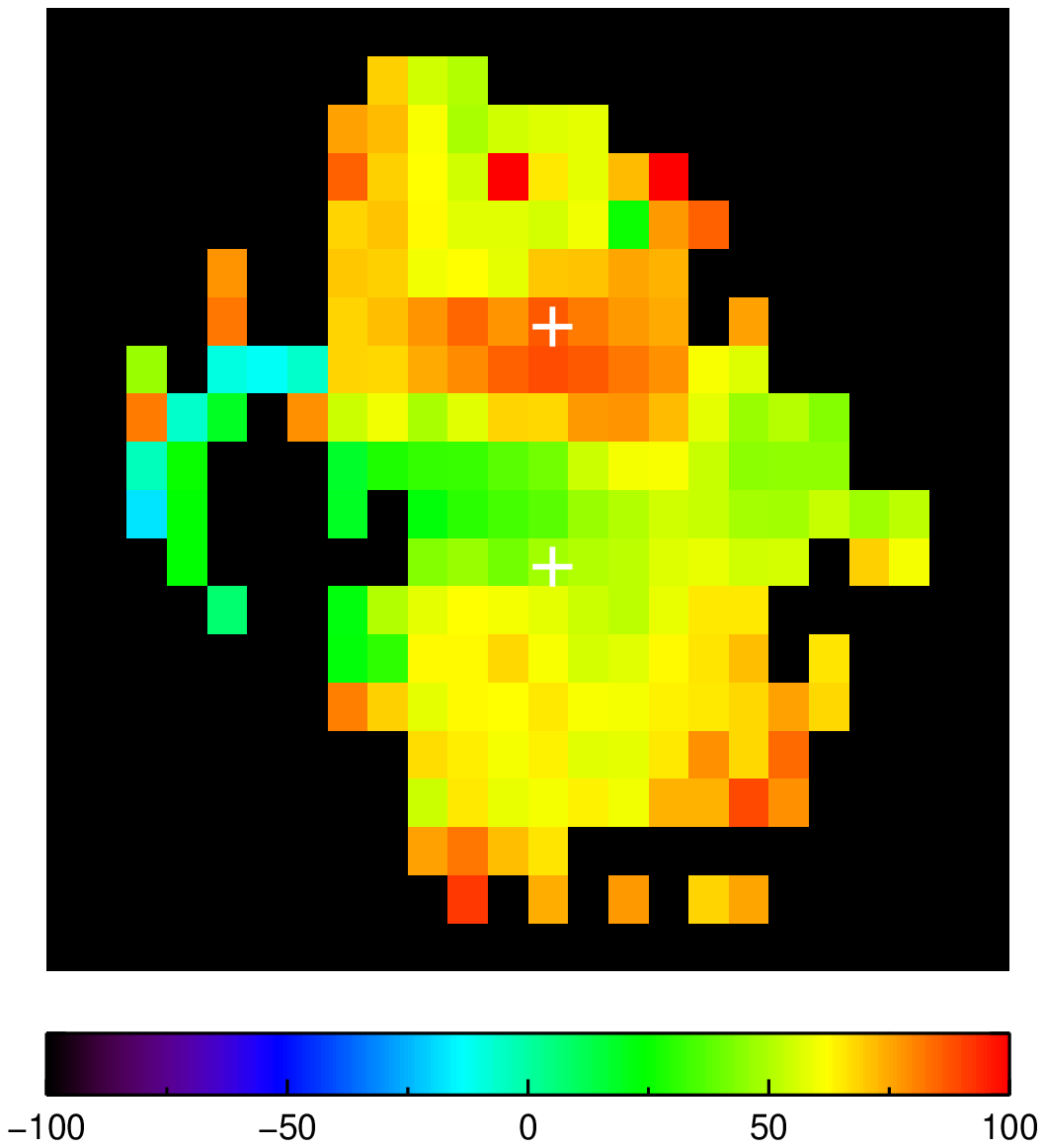,width=4cm}}\quad
\subfigure[Velocity(km s$^{-1}$): $\sigma>155$ km s$
^{-1}$]{\epsfig{figure=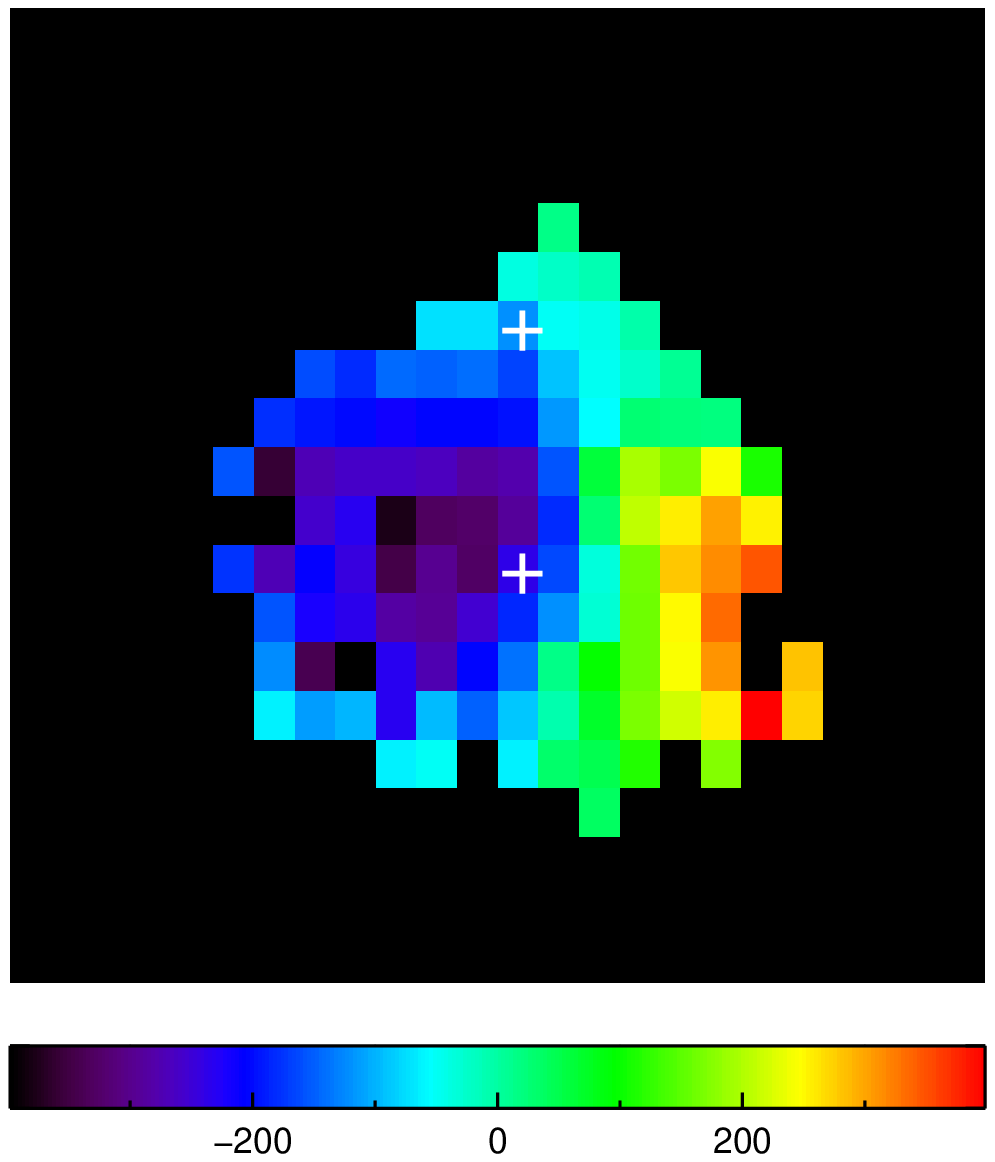,width=4cm}}}
\caption{Maps of velocity for the two components determined by our cutoff. From
left to right the images correspond to the velocity of spaxels with velocity
dispersions of: $\sigma<155$ km s$^{-1}$ and $\sigma>155$ km s$^{-1}$.
Where a spaxel contains two components fit which fall in the same velocity dispersion bin, we show a flux weighted average velocity.
For spatial comparison the HST image has been aligned and placed in the leftmost
panel. The high velocity dispersion gas located to the south shows a velocity shear which we interpret either rotation or outflow.}\label{zmap}
\end{figure*}

Using the separate Gaussian components we are able to calculate the recessional
velocity for each of the two velocity dispersion groups. We adopted a systemic
redshift of z=0.04460 \citep{Lauberts1989}. The velocity maps are shown in Figure
\ref{zmap}. A velocity shear can be seen in the map of the second, broader
component ( $\sigma>155$ km s $^{-1}$), centred over the southern nucleus.
The low velocity dispersion component, which likely
corresponds to star forming regions based on its velocity dispersion, does not
show rotation. However, it is possible that this component could be a face-on
rotating structure, in which case we would not see a velocity shear.

Studying P$\alpha$ in ESO 148-IG002, \cite{PiquerasLopez2012} similarly find low velocity dispersions ($\sim$65 km s$^{-1}$) in the northern nucleus and high velocity dispersion ($\sim$190 km s$^{-1}$) in the southern nucleus. They observe red and blue wings in the P$\alpha$ and H$_2$1-OS(1) line profiles which they suggested forms a cone-like structure, centered on the AGN and extending $\sim 3-4$ kpc north-east and south-west. \cite{PiquerasLopez2012} also report a velocity gradient of $\Delta v\sim$140~km~s$^{-1}$ around the northern nucleus in a north-south direction, a feature which can also be seen in our Figure \ref{vmaps}b. We show, by decomposing emission lines into two Gaussian components, that the velocity shear observed in the south, extends out to $\sim 15$~kpc in the optical.

We propose two possible explanations for the velocity and velocity dispersion fields of ESO 148-IG002.
Firstly, the rotating component seen to the south, could be the disk of the progenitor spiral, containing an AGN.
Due to the large spatial coverage of this rotating component ($\sim 15$~kpc) it is difficult to compare this component to other studies
such as \cite{U2013} who observe rotating gas $<1$~kpc from an AGN. However, larger samples of (U)LIRGs \citep{Medling2014} do see evidence of both small nuclear disks (r $\sim$ few hundred pc) and larger disks (r $>$ 1 kpc), which they also interpret as the progenitor galactic disk.   
The velocity broadening is likely due to the presence of the AGN. However, beam smearing could also cause an increased velocity dispersion. [OIII] is commonly used to measure the size of narrow-line regions (NLRs) in AGNs, typically giving  extents of 1-5 kpc \citep{Bennert2006,Davies2014}. It is unlikely that the AGN at the center of the southern nucleus is able to cause the high line ratios observed ([OIII]/H$\beta$) at 15 kpc without help from star formation or shocks.

Secondly, the velocity shear, which was previously interpreted as rotation, could represent an AGN bipolar outflow, similar to that seen in \cite{Davis2012} and suggested in \cite{PiquerasLopez2012}. \cite{Harrison2012} find broad [OIII] emission lines ($\sigma =$ 300-600 kms$^{-1}$) in high-redshift(z=1.4-3.4) galaxies containing AGN. The broad emission line regions extend across 4-15~kpc and have high velocity offsets from the systemic redshift ($\approx$ 850 km$s^{-1}$) and are attributed to galaxy-scale AGN-driven winds. ESO 148-IG002's broad velocity dispersion gas has a more modest velocity offset ($\sim$350~km~s$^{-1}$) than the outflows in the \cite{Harrison2012} sample, however, the high velocity dispersion, large spatial extent, and velocity shear are consistent with the AGN-driven outflow scenario.

By comparing the ionized gas kinematics with the stellar velocity field, \cite{Davis2012} was able to conclude that their high velocity dispersion component
was due to shocked outflow, as the axis of the gas' velocity gradient was offset from the stellar rotation axis.
Information on the stellar kinematics of ESO 148-IG002's could help differentiate between the two scenarios proposed here.

For the purposes of clarity, we refer to the kinematically distinct broad component as ``rotating'' in subsequent analysis, though the reader should remember that its coherent structure may either be rotation or outflows.

Fitting a disk model to the rotating component could also help decide whether the velocity shear is due to rotation or outflow. Disk fitting and kinemetry analysis \citep{Krajnovic2006} will be subject of future work, with a larger sample of galaxies, allowing us to better understand the kinematic properties of complex galaxy systems.

\section{Analysis}
In the previous section we separated the emission line fluxes into two
components, one with velocity dispersions $\sigma <155$ km s$^{-1}$, and the
other with velocity dispersions higher than $\sigma > 155$ km s$^{-1}$. The
distribution of flux into the two components is illustrated in Figure 8, which
maps the H$\alpha$ flux in the low and high velocity dispersion components. Low
velocity dispersion gas is present throughout the system, whereas the high
velocity dispersion gas is centred on the southern nucleus, where the AGN lies.
In this section we use the emission line ratios of the separate components to
analyse two potentially distinct power sources.

\begin{figure*}
\centering
\mbox{\subfigure[H$\alpha$:
$\sigma<155$km/s]{\epsfig{figure=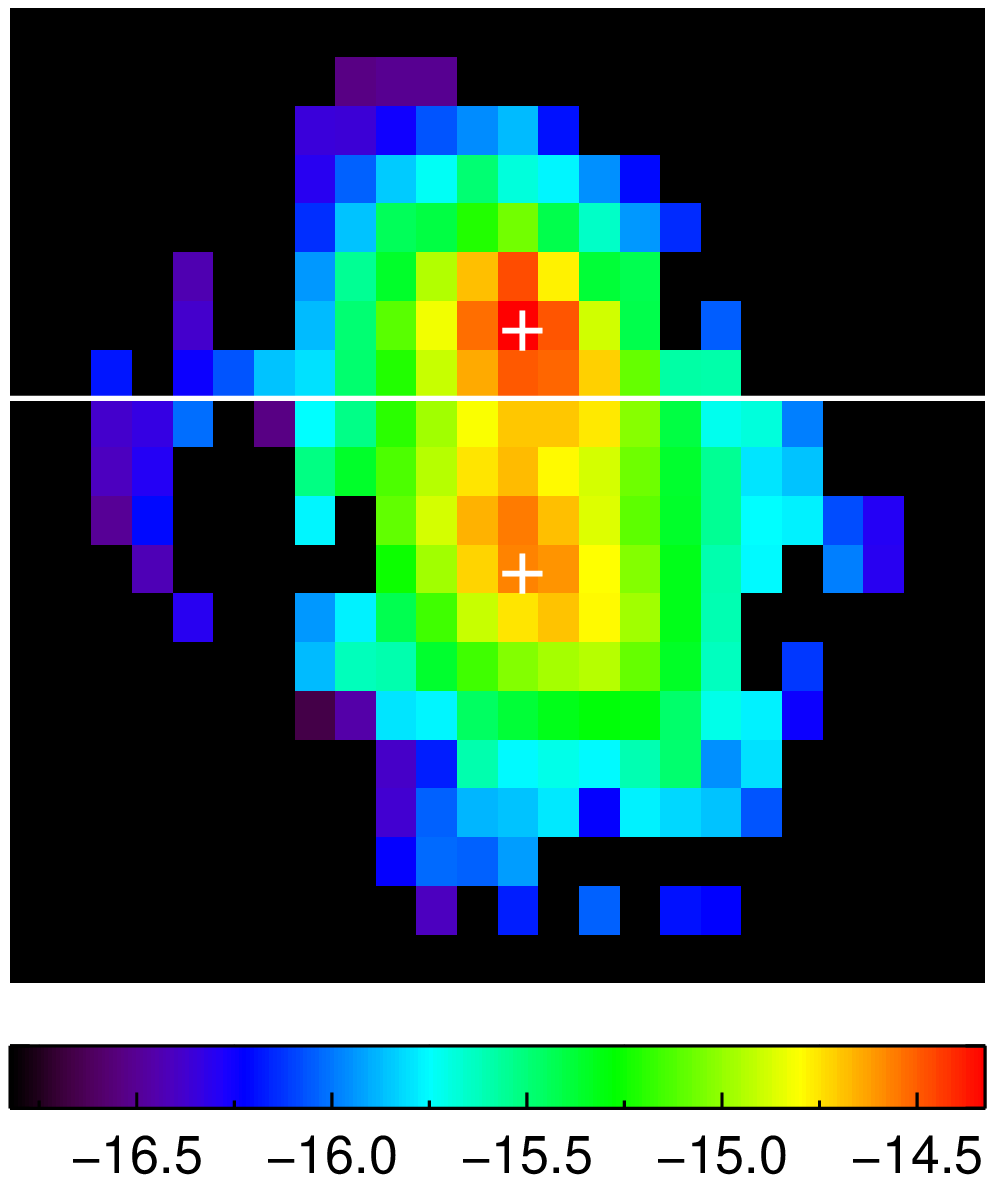,width=3.6cm}}\quad
\subfigure[H$\alpha$ error:
$\sigma<155$km/s]{\epsfig{figure=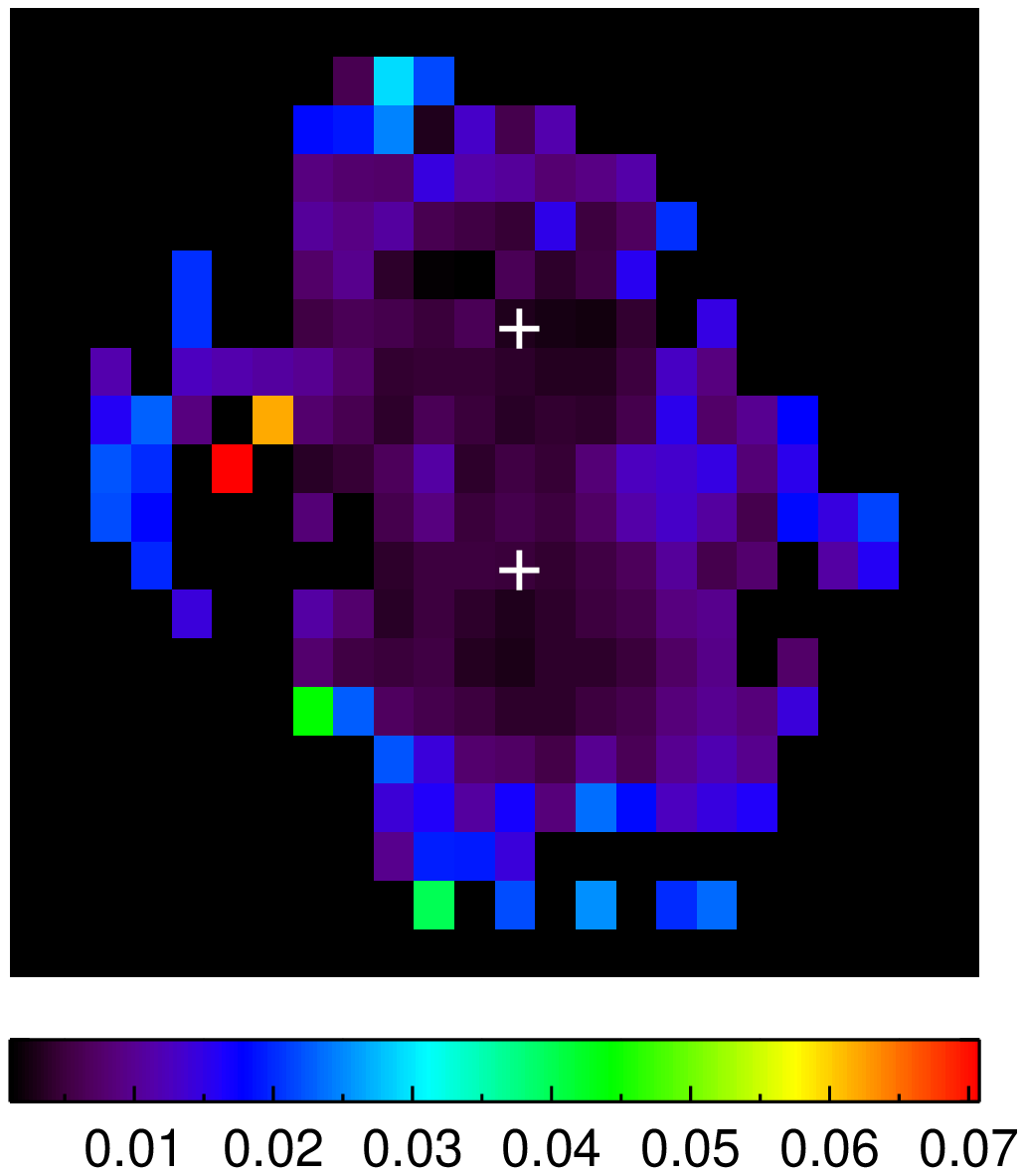,width=3.6cm}}\quad
\subfigure[H$\alpha$:
$\sigma>155$km/s]{\epsfig{figure=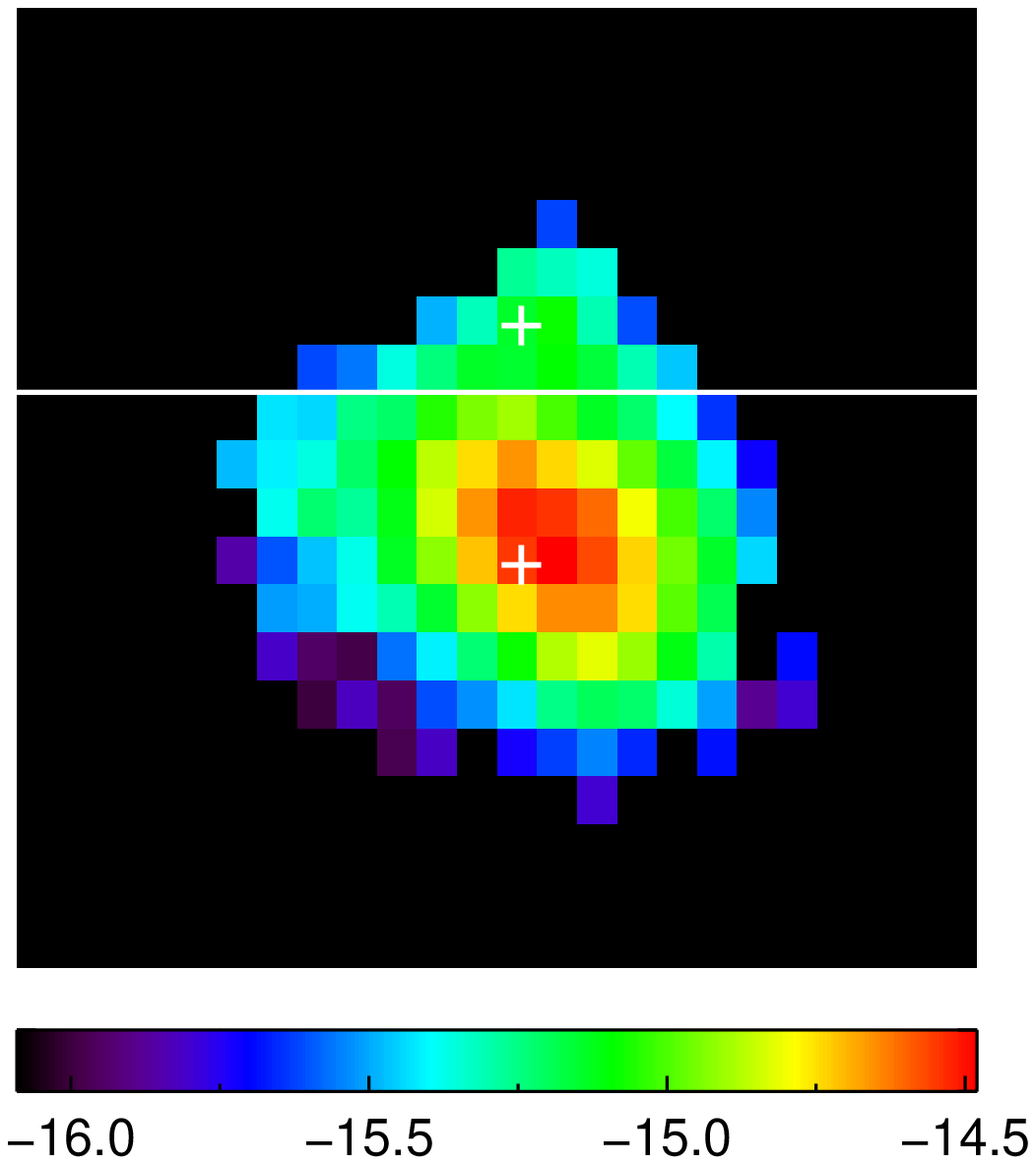,width=3.6cm}}\quad
\subfigure[H$\alpha$ error:
$\sigma>155$km/s]{\epsfig{figure=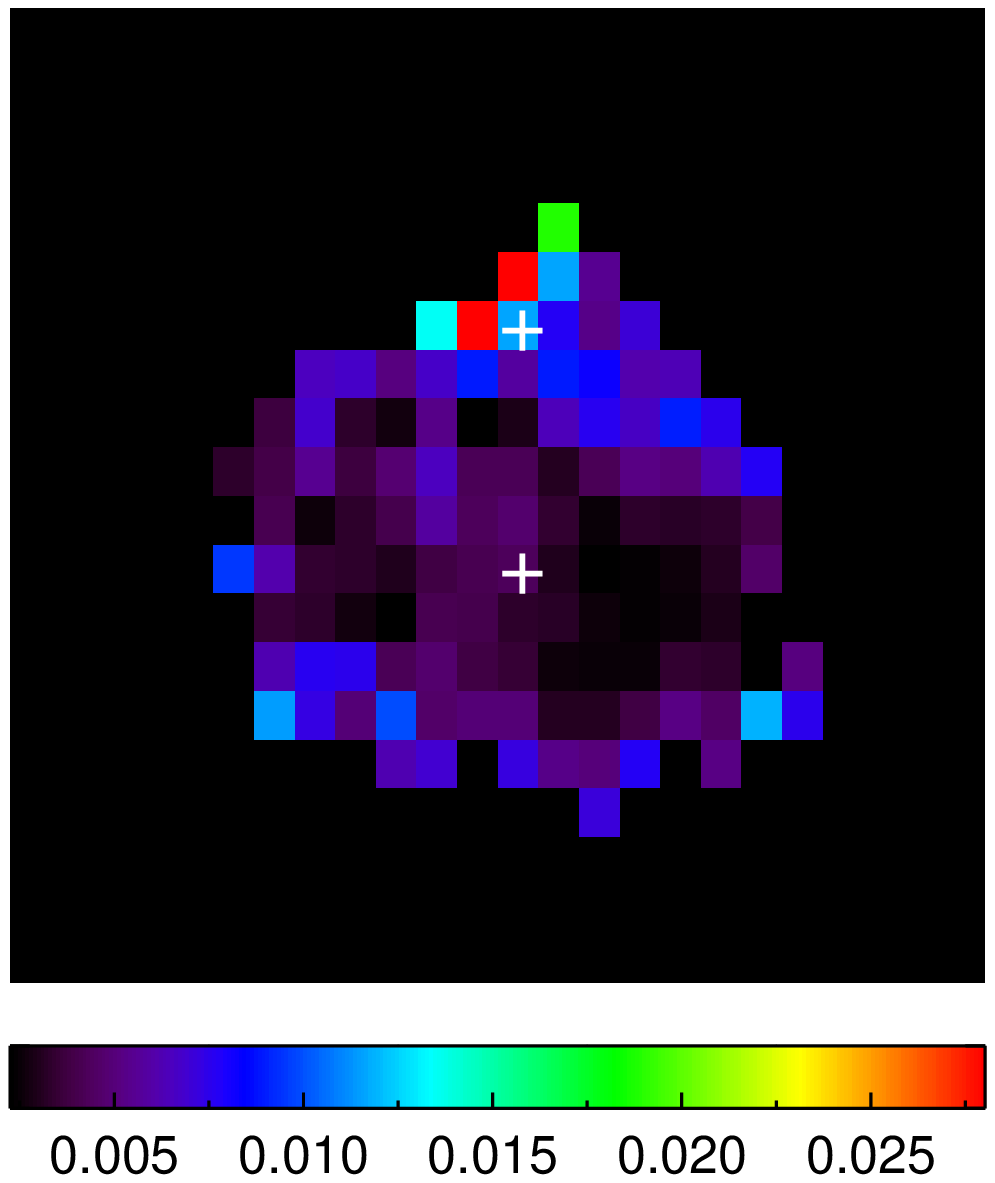,width=3.6cm}}}
\caption{Maps of the log of the H$\alpha$ emission (in units of ergs s$^{-1}$
cm$^{-2}$ \AA$^{-1}$) of the two Gaussian components determined by our velocity
dispersion cutoff and their corresponding uncertainty maps. From left to right
the images correspond to the velocity of spaxels with velocity dispersions of:
$\sigma<155$ km s$^{-1}$ and $\sigma>155$ km $s^{-1}$. The distribution of the two components match with the HST nuclei with the southern component peak
consistent with the AGN position. Horizontal white lines on the flux maps mark the division between the north and south nuclei from which we calculate distances
in Section 5.2.}\label{hacompmap}
\end{figure*}

\subsection{Diagnostic Diagrams}
\begin{figure*}
\centerline{\includegraphics[height=7.5cm]{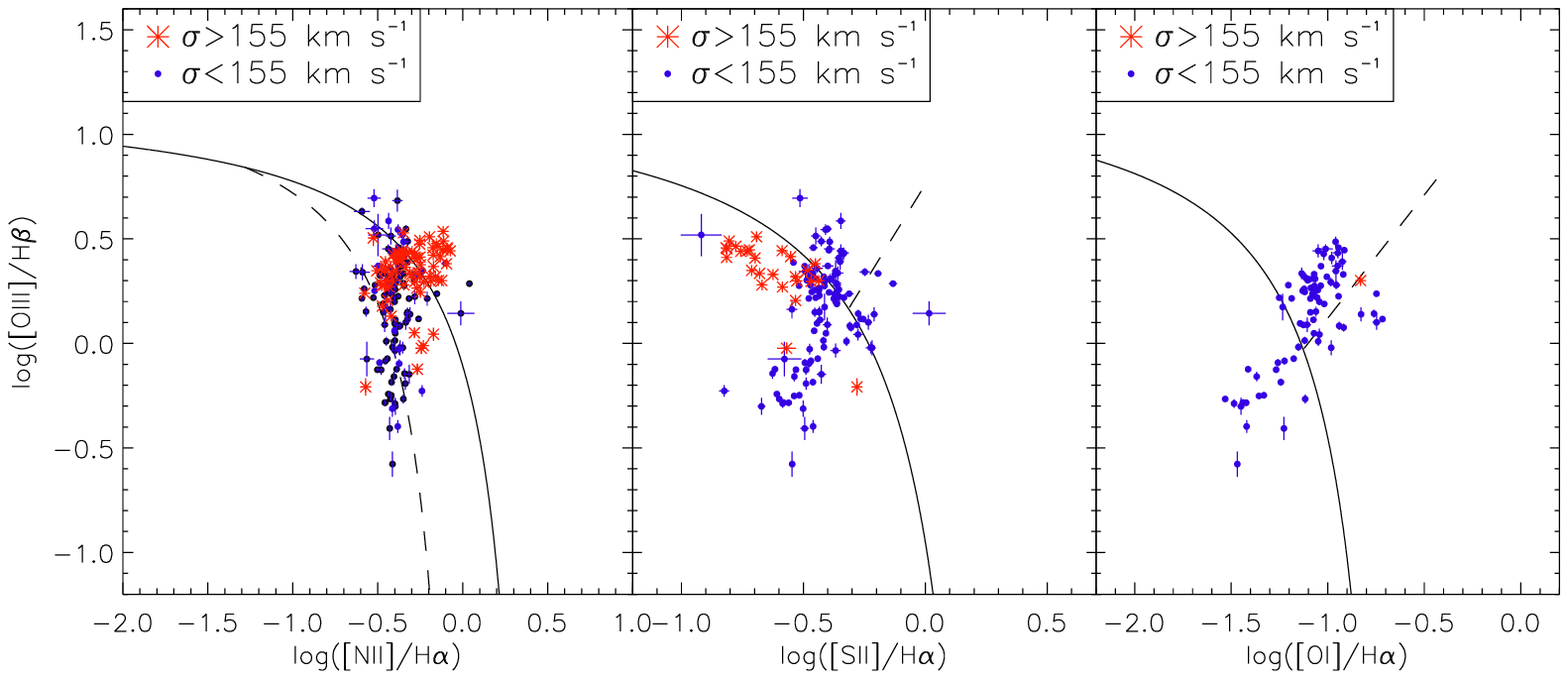}}
\caption{As Figure \ref{BPT}, but with separate Gaussian components color-coded based
on their velocity dispersions. Separating the emission lines in a spaxel in up to two components
results in a greater number of data points than shown in Figure 4. Spaxels with $\sigma$ less
than 155 km s $^{-1}$ are plotted in blue, and the highest velocity dispersions,
$\sigma$ greater than 155 km s$^{-1}$ are coloured red. Every line used has S/N
$> 5$. The broad components are mostly in the Seyfert region on these diagrams and the narrow component spaxels are spread from the star forming region upwards, indicating a larger contribution from star formation in the low velocity dispersion gas.}\label{compBPT}
\end{figure*}

In Figure \ref{compBPT} the spaxels on the optical diagnostic diagrams are
colour coded by velocity dispersion. Line emission from each Gaussian component
of every spectrum with $S/N>5$ is considered here. If the two different colours
occupy separate positions on the diagnostic diagrams, then this separation would
imply that different ionizing sources are responsible for each velocity
component. Separating the emission lines in a spaxel in up to two components
results in a greater number of data points than shown in Figure 4. When two or
more processes are working to power the emission spectrum of the gas, using a
line's total flux for analysis, as in Figure 4, works to average the effects
that the different processes may have. Separating the flux into two separate
components, based on their velocity dispersion, allows us to analyse two
potentially different power sources. There are fewer high velocity dispersion
components present in the [SII]/H$\alpha$ diagnostic diagram and only one in the
[OI]/H$\alpha$ diagram. The lack of high velocity dispersion spaxels is a result
of applying a signal to noise cut on each component fit.\\

Figure \ref{compBPT} shows that the broad component is dominated by spaxels in
the Seyfert and composite region of
the diagnostic diagram, while a larger fraction of the narrow component lies in
the HII-region portion of the diagnostic diagram. The low velocity dispersion
components spread upwards predominantly to the AGN regions in both the [SII] and
[OI] diagnostic diagrams indicating that even the low velocity dispersion gas
could be affected by the AGN and/or shocks.

\subsection{Metallicity Gradient}
\begin{figure}
\centerline{\includegraphics[width=\linewidth]{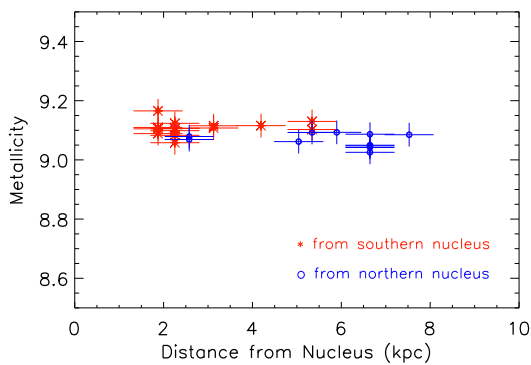}}
\caption{Metallicity versus distance from corresponding nucleus. Metallicities
in red lie to the north of the system. For these metallicities, the distance has
been measured from the northern nucleus. Blue metallicities lie to the south of
the galaxy and the distance to them is calculated from the southern
nucleus. Uncertainties in the distance from the nucleus are derived from the average seeing of our observations. No gradient in the metallicity as a function of radius is evident.}\label{metallicity}
\end{figure}

To find the metallicity (given as 12+log(O/H)) of ESO 148-IG002, we use the method of \cite{Kobulnicky2004},
which uses the stellar evolution and photoionization grids from \cite{Kewley2002} to produce an (updated)
analytic prescription for estimating oxygen abundances using the traditional strong emission line ratio $R_{23}$,
where $R_{23}= ([OII]\lambda 3727/[OIII]\lambda\lambda 4959,5007)/H\beta$.
The $R_{23}$ calibration is sensitive to the ionization state of the gas, particularly for low metallicities.
The ionization state of the gas is characterised by the ionization parameter (the number of hydrogen-ionizing photons passing through
a unit area per second, divided by the hydrogen density of the gas), which is typically derived using the [OIII]/[OII] ratio.
However, this ratio is also sensitive to metallicity, so \cite{Kobulnicky2004} suggest an iterative approach to derive a consistent ionization
and metallicity solution. We first determine whether the spaxels lie on the upper or lower $R_{23}$ branch using the [NII]/[OII] ratio
and calculate an initial ionization parameter assuming a nominal lower branch [12+log(O/H)=8.2] or upper branch [12+log(O/H)=8.7]
metallicity using:
\begin{eqnarray}
\log(q)=(32.81\! -1.153y^2\! +[12+\log(O/H)](-3.396\notag\\
-0.025y+0.1444y^2)(4.603-0.3119y-0.163y^2\notag\\
+[12+\log(O/H)](-0.48+0.0271y+0.02037y^2))^{-1}
\end{eqnarray}
where $y=\log([OIII]\lambda 5007/[OII]\lambda 3727)$.
A spaxel is determined to lie on the lower $R_{23}$ branch if $\log([NII]/[OII])<-1.2$, and on the upper branch if
$\log([NII]/[OII])>-1.2$. The initial ionization parameter is then used to derive a metallicity estimate, using
\begin{eqnarray}
12+\log(O/H)_{lower}=9.40+4.65x-3.17x^2\notag\\
\qquad \quad -\log(q)(0.272+0.547x-0.513x^2)
\end{eqnarray}
where $x=\log(R_{23}),$ if the spaxel lies on the lower branch, and
\begin{eqnarray}
12+\log(O/H)_{upper}=9.72-0.777x-0.951x^2-0.072x^3\notag\\
\quad -0.811x^4-\log(q)(0.0737-0.0713x-0.141x^2\notag\\
\qquad +0.0373x^3-0.058x^4),
\end{eqnarray} if the spaxel lies on the upper branch. Equations 1 and 2 (or 3) are iterated until $12+\log(O/H)$ converges.
To find the metallicity of spaxels dominated by star-formation, emission line
fluxes from the low velocity dispersion component, with S/N greater than 5 in
all relevant lines, lying to the left of the empirical line from \cite{Kauffmann2003} (seen in Figure \ref{compBPT}) were used, and an average metallicity of
$log(\frac{O}{H})+12=  9.09$ ($\pm 0.03$ dex) was obtained.
Metallicity as a function of radius is plotted in Figure \ref{metallicity}. As
there are two nuclei in ESO 148-IG002, we have separated the spaxels into two
groups, a northern and southern, using the spatial cut-off given by the boundary
of the rotating broad component gas (Figure \ref{hacompmap}). Spaxels belonging to the northern nucleus
were determined to be spaxels north of the broad component seen in
Section 4.2, and as such the distance to these spaxels is given from the
northern nucleus. Distances to spaxels south of this spatial boundary are
calculated from the position of the southern nucleus. The gas metallicity of the
star forming regions remains constant around the mean value as a function of
nuclear distance, with a scatter of $\sim 0.2$ dex.

Normal spiral galaxies have negative metallicity gradients consistent with
central enrichment from generations of star formation \citep{Henry1999}. Figure \ref{metallicity}
indicates that ESO 148-IG002 has a flat metallicity gradient. Flat metallicity
gradients can be produced by merger-induced gas inflows
\citep{Kewley2010,Rupke2010,Torrey2012}. The gradient in ESO 148-IG002 is flatter than
any mergers in \cite{Kewley2010} and may indicate very recent gas infall.

\subsection{AGN, Shock and HII models}
\begin{figure*}
\centerline{\includegraphics[height=8cm]{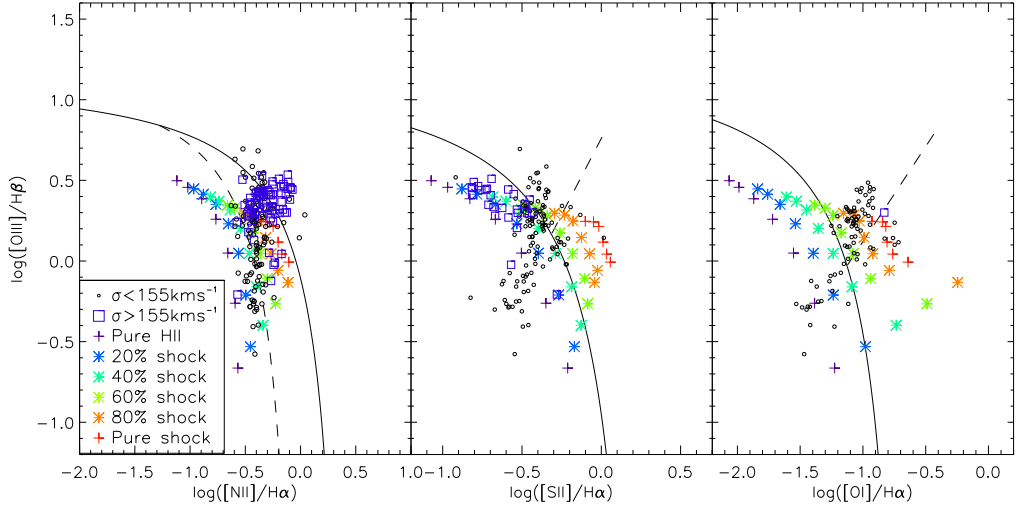}}
\caption{As Figure \ref{compBPT}, but with spaxels corresponding to
the high velocity dispersion gas in blue, and the low velocity dispersion gas in
black. Overplotted are HII region and shock models, with varying ionization
parameter (6.5-8.0), and shock speed (100-200 km s$^{-1}$) for solar metallicity
(8.69). In between, we show a mixing sequence from pure HII region to
pure shock excitation. \textit{Increasing shock velocity} or \textit{decreasing ionization parameter} decreases the [OIII]/H$\beta$ ratio and increases the other ratios. The shock mixing models do cover some of the data but are unable to explain all of ESO 148-IG002's line ratios. Parts of the shock mixing models fall on the same region as AGN mixing models (Figure 12), as they both start from the HII region model. Thus even though some spaxels are consistent with our shock models, we can not conclude that the gas is shocked.}\label{shock}
\end{figure*}

To determine the effects (if any) of shock excitation, we employ slow shock
models, which were introduced and described in \cite{Farage2010} and \cite{Rich2010}. Slow shock models with velocities consistent with our observed line
widths, (100-200 km s$^{-1}$) were generated using an updated version of
Mappings III code, originally introduced in \cite{Sutherland1993}. We
also compare HII region models generated using Starburst99 \citep{Leitherer1999} and
Mappings III \citep{Kewley2001,Levesque2010}. We show mixing sequences of star
formation and shock excitation in Figure \ref{shock} for a metallicity of 8.69
(solar), calculated by varying the fractional contribution of the HII and shock models on the line ratios. Higher metallicity models would lie lower on the diagrams and are
inconsistent with the position of ESO 148-IG002. Not all spaxels are covered by the shock-mixing models shown in Figure \ref{shock}. Therefore although some spaxels may be influenced by shocks, it is likely that (a mixing between star formation and) another phenomenon is causing the high line ratios seen.

\begin{figure*}
\centerline{\includegraphics[height=8cm]{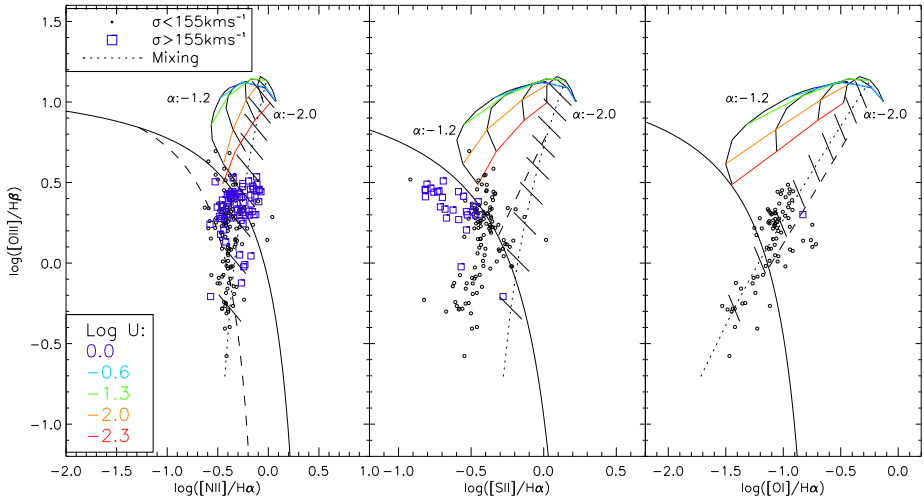}}
\caption{As Figure \ref{compBPT}, but with spaxels corresponding to
the high velocity dispersion gas in blue, and the low velocity dispersion gas in
black. Included are the twice solar metallicity AGN models from \cite{Groves2004}. The
power-law index $\alpha$ values (black grid lines) represented are -1.2, -1.4, -1.7, and -2.0. Ionization parameter ($\log U_0$) values represented include 0.0,
-0.6, -1.3, -2.0, and -2.3. A mixing sequence between a pure HII region and a
pure twice solar metallicity AGN region is created, with each line representing
a change in the fraction of starburst to AGN of 10 $\% $ in linear
space (dotted line). A combination of AGN and HII region models better describe our line ratios than the shock models of Figure \ref{shock}.}\label{agn}
\end{figure*}

We also employ a dusty, radiation pressure-dominated photoionization model first
introduced in \cite{Dopita2002}, and updated in \cite{Groves2004}. These
models provide a self consistent explanation for the emission from narrow-line
regions of AGN. From the family of models available, we explored those with
hydrogen number density of $10^3$ cm$^{-3}$ and a metallicity of 2Z$_\odot$ to
match the properties of ESO 148-IG002. A simple power law represents the
spectral energy ($\nu$) distribution of the ionizing source, with $$ F_\nu
\propto \nu^{\alpha}, \nu_{min}<\nu<\nu_{max}.$$ and $\nu_{min}=$ 5eV and
$\nu_{max}=$ 1000eV. Four values of the  power-law index, $\alpha$ are shown in
Figure \ref{agn}; -1.2,-1.4,-1.7, and -2.0 as these indices encompass the range
of indices seen in AGN locally \citep{Groves2004}. The ionization parameter, log
$U_0$ in the model shown varies between 0.0 and -2.3 dex. We also construct a
starburst-AGN mixing sequence, by varying the fractional contribution of the HII and AGN models on the line ratios in linear space. The mixing sequence shown in Figure \ref{agn} results from applying this method on a single pure HII region and pure twice solar metallicity AGN region indicating changes in the fraction of starburst to AGN of 10 $\% $ in linear space.

Neither model encompasses all spaxels from ESO 148-IG002. Recall from Figure
\ref{compBPT} that almost all of the spaxels on the [OI] diagnostic are of lower
velocity dispersion. Low velocity dispersion spaxels lie on both the shocked mixing sequence and the AGN mixing sequence. Spaxels in the higher velocity dispersion
group are closer to the AGN models, which indicates that this high
$\sigma$ component is influenced more strongly by the AGN in the southern
nucleus, over which those spaxels lie.

To determine whether the starburst-AGN or starburst-shock models better describes our emission line data, we perform the statistical Kolmogorov-Smirnov (KS) test. The KS test is a nonparametric test of the equality of two continuous distributions and can be used to determine the likelihood that a sample distribution was drawn from reference distribution (one-sided KS test), or to compare two sample distributions (two-sided KS test).
We compare the values of ESO 148-IG002's [OIII]/H$\beta$ and [SII]/H$\alpha$ ratios with the values estimated from the above mixing sequences using a two-dimensional, two-sided KS test. Although the KS test is only strictly defined for one dimensional probability distributions, \cite{Press2002} describes how an analogous test can be designed and implemented when the distribution depends on two variables. We use the combination of [OIII]/H$\beta$ and [SII]/H$\alpha$ ratios for this test as the [SII]/H$\alpha$ ratio is more sensitive to shocks than [NII]/H$\alpha$, and is of higher signal to noise in our data than the [OI]/H$\alpha$. Table 1. lists the $p$ values calculated using the two sided KS test. A $p$ value close to one means that there is a high probability that the two samples originated from the same parent distribution. We find that gas with $\sigma<155$ km s$^{-1}$ can be better described by the AGN mixing sequence, than with the shock mixing sequence used, with $p$ values 0.73 and 0.59 respectively. The high velocity dispersion gas with $\sigma>155$ km s$^{-1}$ is also better described by our AGN mixing model, with $p$=0.63 as compared with $p$=0.48 for the shock mixing model. As previous studies have indicated the presence of an AGN in the southern nucleus as well as star formation, it is not surprising that both the line ratios and velocity dispersions of ESO 148-IG02 show a mixture of star formation and AGN activity. Both the low and high velocity dispersion components are better described by the model AGN mixing sequence than the shock mixing sequence. It is likely that the high velocity dispersion component traces gas strongly influenced by the AGN, whilst the low velocity dispersion gas is more strongly influenced by star formation with AGN activity having a small effect.

Figure 11 suggests that more theoretical work is needed to accurately model the [SII] emission-line fluxes in merging galaxies.  Unlike the [NII]/Ha and [OI]/Ha mixing sequences, the [SII]/Ha line ratios are not well-fit by the mixing sequence.  The difficulty in modeling the [SII]/Ha lines in galaxies is well known (e.g., \citealt{Levesque2010}), and has been attributed to
the theoretical shape of the EUV radiation field.

\begin{table}
\centering
\begin{tabular}{ccc}
\hline
$p$ value & shock & AGN \\
\hline
$\sigma<155 km/s$ & 0.587 & 0.729\\
$\sigma>155 km/s$ & 0.481 & 0.628\\
\hline
\end{tabular}
\caption{The results of performing the KS test which compared the ratios of [SII]/H$\alpha$ and [OIII]/H$\beta$ for each spaxel, with a model star forming-shock mixing sequence, and a model star forming- AGN mixing sequence. Values of $p$ closer to one imply that the underlying distributions are more likely the same, i.e that the model is preferred. We find that the AGN mixing sequence better describes our data than the shock mixing sequence.}
\end{table}

\section{Discussion}
Observations by \cite{Rich2012} and \cite{Kewley2010}, and simulations
\citep{Rupke2010} show that there is a strong relationship between metallicity
gradients and the gas dynamics in galaxy interactions and mergers. As a merger
progresses the metallicity gradient is flattened. This flattening reflects the
effects of gas redistribution over the disks, including less enriched gas being
torqued towards the centre and the growth of tidal tails that carry metals out
to large radius. In the previous section we found the metallicity gradient of
ESO 148-IG002 is flatter than any of the sample galaxies in \citet{Kewley2010}.
Based on our analysis, we present the following picture of ESO 148-IG002:

Gas from the northern galaxy has been spread out across the whole system as a
result of the merger. This constitutes the low velocity dispersion component
($\sigma <155 $km s$^{-1}$) which covers ESO 148-IG002. This area is most likely
to be dominated by star formation as indicated from the optical diagnostic line
ratios. The mixing of the gas due to the gravitational forces acting between the
two galaxies is likely to be responsible for the flat metallicity gradient
observed.
The southern galaxy seems to have remained dynamically distinct from the
northern galaxy, either maintaining a rotation that is not seen in the gas with lower
velocity dispersion, or being the result of an AGN-driven outflow. The southern nucleus contains an AGN which is the most
likely cause for the high velocity dispersion in the surrounding gas ($\sigma >
155$km s $^{-1}$). The few spaxels that lie near the southern nucleus have
velocity dispersions of $\sim$600 km s$^{-1}$, consistent with an AGN. It is
not possible to entirely rule out shocks from the galaxy. From the line ratios it
is possible that at least some of the galaxy is influenced by shock excitation.
\\ Considering this interesting scenario and the flat metallicity gradient of
the galaxy, we suggest that we are observing ESO 148-IG002 in the middle of (or
just after) major gas rearrangement related to its merger, which makes it an
exciting candidate for future studies of star formation and AGN fuelling.

\section{Conclusion}
Using WiFeS wide integral field spectroscopy of the ULIRG ESO 148-IG 02 we found
that:\\ \\
$\bullet$ The distribution of gas velocity dispersion is bimodal, indicating a
combination of power sources.\\
$\bullet$ The emission line ratios indicate composite starburst and AGN
activity, with a starburst-AGN mixing model better able to explain the data than a starburst-shock mixing model (AGN mixing model produced $p$ values of 0.73 and 0.63 for the low and high velocity dispersion gas respectively, compared to $p$ values of 0.59 and 0.48 for shock mixing models when tested using the KS statistic).\\
$\bullet$ The HII regions, dominated by star formation and given by the
lower-$\sigma$ component, have a mean metallicity of 9.09$\pm$0.03.\\
$\bullet$ The galaxy has a flat metallicity gradient as a result of the merging
process.\\
$\bullet$ The high $\sigma$ gas associated with the southern nucleus has a coherent velocity pattern which could either be rotation or an AGN-driven outflow.\\

It is possible that we have caught this merger as the gas from the northern
galaxy is being mixed throughout the two galaxies. If this is the case, we have
shown that it is possible to disentangle two galaxies in a merging system based
entirely on its kinematic analysis. \\ ESO 148-IG002 provides an exciting test
bed for future work on the potential triggering of starbursts and AGN by merger
induced gas flows.\\ \\
\textbf{ACKNOWLEDGMENTS}\\ \\
L.K. gratefully acknowledges the support of an ARC Future Fellowship and ARC
Discovery Project grant DP130103925.
We would like to thank A. Medling for discussing the interpretation of the kinematic components with us.
This research has made use of NASA's Astrophysics Data System Bibliographic
Services and the NASA/IPAC Extragalactic Database (NED). The authors are very
grateful to two anonymous referees whose comments and suggestions greatly improved
the clarity of this paper.
\bibliographystyle{aa}
\bibliography{bibliography}{99}
\end{document}